\documentclass[11pt,a4paper,nofootinbib,superscriptaddress]{revtex4-1}
\pdfoutput=1
\usepackage[normalem]{ulem}
\usepackage[utf8]{inputenc}
\usepackage{graphicx,bm}
\usepackage{slashed,verbatim}
\usepackage{amssymb,graphicx,epstopdf}

 \usepackage[colorlinks=true,linktocpage=true,linkcolor=blue,citecolor=blue]{hyperref}
\usepackage{slashed,subfigure}
\usepackage{caption}
\usepackage{amsmath,amssymb,graphicx,xcolor,mathtools}
\usepackage{epstopdf,dcolumn}
\usepackage{soul}

\allowdisplaybreaks
\usepackage[normalem]{ulem}
\usepackage{cancel}
\usepackage{xcolor}
\setstcolor{red}

\def\be {\begin{equation}}
\def\ee {\end{equation}}
\def\bea {\begin{eqnarray}}
\def\eea {\end{eqnarray}}
\def\bc {\begin{center}}
\def\ec {\end{center}}
\def\nn {\nonumber}

\def\mn {\mu\nu}
\def\mr {\mu\rho}
\def\rn {\rho\nu}

\def\sp{\shortparallel}

\def\({\left(}
\def\){\right)}
\def\[{\left[}
\def\]{\right]}

\def\sumintb{\sum\!\!\!\!\!\!\!\!\!\int\limits}
\def\sumintf{\sum\!\!\!\!\!\!\!\!\!\!\int\limits}

\begin{document}
\title{General structure of gauge boson propagator and its spectra in a hot magnetized medium }
 \author{Bithika Karmakar}
\email{bithika.karmakar@saha.ac.in}
  \affiliation{
 	Theory Division, Saha Institute of Nuclear Physics, HBNI, \\
 	1/AF, Bidhannagar, Kolkata 700064, India. }
 \author{Aritra Bandyopadhyay}
  \email{aritrabanerjee.444@gmail.com}
 \affiliation{Departamento de F\'{\i}sica, Universidade Federal de Santa Maria, 
 	Santa Maria, RS 97105-900, Brazil}
 \author{Najmul Haque}
 \email{nhaque@niser.ac.in}
 \affiliation{School of Physical Sciences, National Institute of Science Education and Research, HBNI,\\  Jatni, Khurda 752050, India}
 \affiliation{Institut f\"ur Theoretische Physik, Justus-Liebig--Universit\"at Giessen, 35392 Giessen, Germany}
 \author{Munshi G Mustafa}
 \email{munshigolam.mustafa@saha.ac.in}
  \affiliation{
 	Theory Division, Saha Institute of Nuclear Physics, HBNI, \\
 	1/AF, Bidhannagar, Kolkata 700064, India. }

\begin{abstract}
{ Based on transversality condition of gauge boson self-energy 
we have systematically constructed the general structure of the gauge boson two-point functions using four linearly independent basis tensors in 
presence of a nontrivial background, {\it{i.e.}}, hot magnetized material medium. The hard thermal loop approximation 
has been used for the  heat bath to compute various form factors associated with the
gauge boson's  two point functions both in strong and weak field approximation. We have also  analyzed the dispersion  
of a gauge boson (e.g., gluon)  using the effective propagator both in  strong  and weak magnetic field
approximation.  The formalism is also applicable to  QED.   
The presence of only thermal  background  
leads to a longitudinal (plasmon) mode and a two-fold degenerate transverse mode.  In presence of a hot  magnetized 
background  medium the degeneracy of the two transverse modes is lifted and one gets three quasiparticle modes. 
In weak field approximation one gets two transverse modes and one plasmon mode. On the other hand,  in strong 
field approximation also one gets the three modes in Lowest Landau Level.
The general structure  of two-point function may be useful for computing the thermo-magnetic correction  
of various quantities associated with a gauge boson.}
\end{abstract}
\date{\today}
\maketitle 

\section{Introduction}
The propagation of vector gauge bosons in a material medium in presence of a magnetic field produces many interesting observational effects.  
 As for example  the  photons with different polarizations have 
different dispersion properties which lead  to the Faraday rotation. This has also been observed for various astrophysical 
objects~\cite{Giovannini:1997km,Giovannini:1997eg, Kronberg:1993vk, Kosowsky:1996yc} and in the millisecond pulsations 
of solar radio emission\cite{Fleishman:2002zz}. 
In view of the theoretical perspective  the general feature is  associated with the propagation of a photon in an externally magnetized 
material medium. The subject of the propagation of photons in magnetized plasmas  has been studied in large extent and also 
covered in standard  electromagnetic theory~\cite{portis1978,rai1999} and plasma physics~\cite{liftshitz1981, ishimaru1992} books. 
However, in most cases it was assumed that the medium consists of non-relativistic and non-degenerate 
electrons and nucleons.  This suggests a modification of  theoretical tools in which a  general formalism based on 
quantum field theory proves to be helpful \cite{Nieves:1988qz}. A quantum field theoretical formalism to calculate 
Faraday rotation in different kinds of media (hot magnetized one) have been done in Refs.~\cite{Ganguly:1999ts,DOlivo:2002omk}.
Also high-intensity laser fields  are used to create  ultrarelativistic electron-positron plasmas  which  play an important 
role in various astrophysical situations. Some properties of such plasma are studied  using QED at finite temperature~\cite{Thoma:2008my,Thoma:2008gh}.

In the regime of Quantum Chromo Dynamics (QCD), nuclear matter dissolves into a thermalized color deconfined state Quark Gluon Plasma (QGP) 
under extreme conditions such as very high temperature and/or density. To probe different characteristics of this novel state, various high energy 
Heavy-Ion-Collisions (HIC) experiments are under way, \textit{e.g}, RHIC@BNL, LHC@CERN and  upcoming  FAIR@GSI. Depending on the 
impact parameter of the collision, a relativistic HIC can be central or non-central. In recent years the non-central HIC is getting more and more attention 
in the heavy-ion community because of some distinct features which appear due to the non-centrality of the collision. One of those is the  
prospect of producing a very strong magnetic field in the direction perpendicular to the reaction plane due to the relatively higher rapidity of the spectator 
particles that are not participating in the collisions. Presently immense  activities are in progress to study the properties of strongly interacting matter 
in presence of an external magnetic field, resulting in the emergence of several novel 
phenomena~\cite{Shovkovy:2012zn,D'Elia:2012tr,Fukushima:2012vr,Mueller:2014tea,
Miransky:2015ava,Bzdak:2012fr, Basar:2012bp,Fukushima:2008xe, Alexandre:2000yf,
Gusynin:1997kj, Lee:1997zj, Kharzeev:2009fn, Bali:2011qj, Farias:2016gmy,
Ayala:2014iba, Ayala:2014gwa,Ayala:2015bgv, Mueller:2015fka}. 
This suggests that  there is clearly an increasing demand to study the effects of intense background magnetic fields on various aspects and observables  
of non-central heavy-ion collisions. Also experimental evidences of photon anisotropy, provided by the PHENIX Collaboration~\cite{Adare:2011zr}, have posed a challenge for existing theoretical models. This kind of current experimental evidences have prompted that a modification of the present theoretical tools are much needed by considering the effects of intense background magnetic field on various aspects and observables of non-central HIC. In a field theoretic calculation $n$-point functions are the basic quantities to compute the various observables of a system.  With this perspective   in  very recent works,
based on various symmetries of the system for a nontrivial medium like a hot magnetized  one, the general structure of fermionic 2-  and  3-point function~\cite{Das:2017vfh}, and 4-point function~\cite{Haque:2017nxq} were computed. Also the spectral representation of two point function~\cite{Das:2017vfh}  
were obtained for such system. In this paper we consider gluon that propagates in a hot magnetized QCD plasma for which
we  aim  at the general structure of the gauge boson self-energy, the effective propagator and its dispersion property. This formalism is also applicable to QED system. The general propagators for fermion obtained in Ref.~\cite{Das:2017vfh} and for the gauge boson obtained here have already been used to compute the quark-gluon 
free energy for a hot magnetized deconfined QCD system in Ref.~\cite{Bandyopadhyay:2017cle}.

This paper has been organized as follows: in section~\ref{gsse} the general structure of a gauge boson self-energy in 
a hot magnetized medium is discussed progressively.  It includes  two parts: a brief review of the general structure in  presence of  only  thermal 
medium in subsection~\ref{ftb0}  and then a generalization of it to a hot magnetized medium in subsection~\ref{ftb}.  In section~\ref{gsprop} we discuss the general structure 
for the gauge boson propagator using the results of section \ref{gsse}. Section~\ref{ff} begins with  the domain of applicability depending upon  the scales (mass, temperature and 
the magnetic field strength) associated with the system. In subsections~\ref{sfa} and~\ref{wfa} we elaborately compute the  various 
form factors, Debye screening mass, dispersion relations within strong and weak field approximation, respectively. Finally, we conclude in section~\ref{conclusion}.

\section{General Structure of a gauge boson self-energy}
\label{gsse}
In this section we first briefly review the formalism of the general structure for a gauge boson self-energy by considering 
only thermal bath without the presence of any magnetic field  in subsection~\ref{ftb0} and it will  then be followed by a formalism for 
a magnetized hot medium in subsection~\ref{ftb}. 

\subsection{Finite temperature and zero magnetic field case}
\label{ftb0}

We begin with the general structure of the gauge boson self-energy in vacuum, given as
\bea
\Pi^{\mn}(P) = V^{\mn}\Pi(P^2),
\eea
where the form factor  $\Pi(P^2)$ is  Lorentz invariant and 
depends only on the four scalar $P^2$. The vacuum projection operator is
\bea
V^{\mu\nu} = g^{\mu\nu}-\frac{P^\mu P^\nu}{P^2},
\eea
with the metric $g^{\mu\nu}\equiv(1,-1,-1,-1)$ and $P^\mu\equiv(p_0, \bm{p})=(p^0,p^1,p^2,p^3)$.  
The self-energy satisfies the gauge invariance through the transversality condition 
\bea
P_\mu \Pi^{\mu\nu} (P)= 0 ,
\label{trans_cond}
\eea
and it is also symmetric 
\bea
\Pi^{\mu\nu} (P)= \Pi^{\nu\mu} (P).
\label{symm_cond}
\eea
The conditions in Eqs.~(\ref{trans_cond}) and (\ref{symm_cond}) are sufficient to obtain ten components of $\Pi^{\mu\nu}$.

The presence of finite temperature ($\beta=1/T$) or heat bath breaks the Lorentz (boost) invariance
of the system.  In finite temperature one accumulates  four vectors and tensors 
to form a general  covariant structure of the gauge boson self-energy.  Those are 
$P^\mu,  \, g^{\mu\nu}$ from vacuum and the four-velocity $u^\mu$ of the heat bath, discreetly introduced because 
of the medium. With these one can form four symmetric basis tensors, 
namely $P^\mu P^\nu, P^\mu u^\nu + u^\mu P^\nu, u^\mu u^\nu$ and $g^{\mu\nu}$. These four tensors can 
be reduced to two independent mutually orthogonal projection tensors by virtue of the constraints provided by
 the transversality condition  in Eq.~(\ref{trans_cond}). One uses them  to construct  manifestly Lorentz-invariant 
 structure of the gauge boson self-energy and propagator 
at finite temperature which have been discussed in the literature in 
details~\cite{ Das;1997,  Lebellac;1996, Kapusta:1989tk}. Nevertheless,  
we briefly discuss some of the essential points  that would be very useful in constructing those general structures for a magnetized hot medium.
 
We now begin by defining Lorentz scalars, vectors and tensors that characterize 
the heat bath or hot medium in a local rest frame:
 \begin{subequations}
\begin{align}
u^\mu &=(1,0,0,0), \nn\\
P^\mu u_\mu&=P\cdot u=p_0 ,\label{scal1}\\
\tilde{P}^\mu &= P^\mu - (P\cdot u)u^\mu 
= P^\mu - p_0 u^\mu,\\
\tilde{g}^{\mu\nu} &= g^{\mu\nu} - u^\mu u^\nu \\
\tilde{P}^2 &= \tilde{P}^\mu \tilde{P}_\mu = P^2-p_0^2= -p^2, \label{scal2}
\end{align}
\end{subequations}
where $p=|\bm{p}|$. We note here that one can only construct two independent Lorentz scalars as given in Eqs.~(\ref{scal1}) and~(\ref{scal2}).
One can further redefine four vector $u^\mu$ by projecting the vacuum projection 
tensor upon it as 
\bea
\bar{u}^\mu &=& V^{\mu\nu}u_\nu = u^\mu - \frac{(P\cdot u)P^\mu}{P^2} 
= u^\mu - \frac{p_0 P^\mu}{P^2}.
\eea
which is orthogonal to $P^\mu$.
Now one can construct two independent and mutually transverse second rank projection 
tensors in terms of  those redefined set of four-vectors and tensor as
 \begin{subequations}
\begin{align}
A^{\mu\nu} &= \tilde{g}^{\mu\nu} - \frac{\tilde{P}^\mu \tilde{P}^\nu}{\tilde{P}^2}, \label{amunu}\\
B^{\mu\nu} &= \frac{1}{\bar{u}^2} \bar{u}^\mu\bar{u}^\nu . \label{bmunu}
\end{align}
\end{subequations}
Moreover, sum of these two projection operators lead to 
the well known vacuum projection tensor $V^{\mn}$ as 
\bea
A^{\mu\nu} + B^{\mu\nu} &=& g^{\mu\nu} - \frac{P^\mu P^\nu}{P^2} = V^{\mn}.
\eea
So, the general (manifestly) covariant form of the self-energy tensor can be written as
\bea
\Pi^{\mu\nu} = \Pi_T A^{\mu\nu} + \Pi_L B^{\mu\nu},
\label{gen_struc_t}
\eea
where $\Pi_{L}$ and $\Pi_T$ are, respectively, the longitudinal and transverse form factors.
Eventually one can obtain these two form factors as
 \begin{subequations}
\begin{align}
\Pi_L &= -\frac{P^2}{p^2} \Pi_{00},\\
\Pi_T &= \frac{1}{D-2}\left(\Pi^\mu_\mu- \Pi_L\right),
\end{align}
\end{subequations}
where $D$ is the space-time dimension of a given theory. 
The above Lorentz-invariant form factors would depend on the two independent
Lorentz scalars $p_0$ and $ p=\sqrt{p_0^2-P^2}$ as defined, respectively, in 
Eqs.~(\ref{scal1}) and~(\ref{scal2}) besides the temperature $T=1/\beta$.

\subsection{Finite temperature and finite magnetic field case}
\label{ftb}
The finite temperature breaks the Lorentz (boost) symmetry whereas the presence of magnetic field breaks the rotational symmetry in the system. 
In presence of both finite temperature ($\beta=1/T$) and finite magnetic field $B$, the four-vectors and tensors available to form the general covariant structure of the gauge boson self-energy are $P^\mu$, $g^{\mu\nu}$, the electromagnetic field tensor $F^{\mn}$ and it's dual $\tilde F^{\mn}$, and the four velocity of the heat bath,  $u^\mu$.
As seen in section~\ref{ftb0} at finite $T$ the heat bath introduces a preferred direction that breaks the boost invariance. On the other hand, the presence of  
magnetic field breaks  the rotational 
symmetry in the system because it introduces an anisotropy in space. For hot magnetized system, one can define a new four vector
$n^\mu$ which is associated with the electromagnetic field tensor $F^{\mn}$. We define the electromagnetic field tensor as
\bea
   F^{\mn}=
  \left( {\begin{array}{cccc}
   0 & 0 & 0 & 0 \\
   0 & 0 & -B & 0 \\
   0 & B & 0 & 0 \\
    0 & 0 & 0 & 0 \\
  \end{array} } \right).
\eea
In the rest frame of the heat bath, 
{\it i.e.}, $u^\mu =(1,0,0,0)$, $n^\mu$ can  be 
defined uniquely as projection of $F^{\mn}$ along $u^\mu$,
\be
 n_\mu \equiv \frac{1}{2B} \epsilon_{\mu\nu\rho\lambda}\, u^\nu F^{\rho\lambda} 
 = \frac{1}{B}u^\nu {\tilde F}_{\mu\nu} = (0,0,0,1), \label{bmu}
 \ee
which is in the $z$-direction. This also establishes a connection between the heat bath and the magnetic field. 

Now for a hot magnetized case one has Lorentz vectors, $P^\mu$,
 $u^\mu$ and $n^\mu$ along with metric tensor $g^{\mn}$,  from which one can form seven 
symmetric basis tensors, namely $P^\mu P^\nu, P^\mu n^\nu + n^\mu P^\nu, n^\mu n^\nu$, $P^\mu 
u^\nu + u^\mu P^\nu$, $u^\mu u^\nu$, $u^\mu n^\nu + n^\mu u^\nu$ and $g^{\mu\nu}$. These seven tensors reduce to four
because of constraints provided by the gauge invariance condition in Eq.~(\ref{trans_cond}). Below we obtain the four basis tensors\footnote{We note here that a set of
four different basis tensors were used in Refs.~~\cite{Nopoush:2017zbu,Hattori:2017xoo,Ayala:2018wux}.}.

We first form the transverse four momentum and the transverse metric tensor as  
\begin{subequations}
\begin{align}
P_\perp^\mu &=P^\mu - (P\cdot u)u^\mu + (P\cdot n)n^\mu \nn\\
&= P^\mu - p_0 u^\mu + p^3 n^\mu = P^\mu-P_\sp^\mu,\\
g_\perp^{\mn} &= g^{\mu\nu} - u^\mu u^\nu + n^\mu n^\nu = g^{\mu\nu} - g_\sp^{\mu\nu},
\end{align}
\end{subequations}
where 
\begin{subequations}
\begin{align}
P_\sp^\mu &=  p_0 u^\mu -p^3 n^\mu, \label{p_sp}\\
P_\sp^2 &= P_\sp^\mu P^\sp_\mu = p_0^2-p_3^2,\\
g_\sp^{\mu\nu} &= u^\mu u^\nu - n^\mu n^\nu,\label{eta_sp}\\
P_\perp^\mu P^\perp_{\mu} &=P_\perp^2= P^2-p_0^2+p_3^2=P^2-P_\sp^2=-p_\perp^2,
\end{align}
\end{subequations}
where $P^2=P_\sp^2+P_\perp^2=P_\sp^2-p_\perp^2$, $P_\sp^2=p_0^2-p_3^2$ and $p_\perp^2=p_1^2+p_2^2$. 
We further note  that the three independent Lorentz scalars are  $p_0$, $p^3=P\cdot n$  and $P_\perp^2$.

We take $B^{\mn}$ in Eq.~\eqref{bmunu} as one of projection tensors in hot magnetized system.
Now $A^{\mn}A_{\mn}=2$ indicates that it is a combination of two mutually orthogonal projection tensors, which yields
two degenerate transverse modes for gauge boson in heat bath. Projection of $A^{\mn}$ along magnetic field direction $n^{\mu}$
is $\bar n^{\mu}=A^{\mn}n_{\nu}$. So we can construct another second rank  
tensor orthogonal to both $P^\mu$ and $B^{\mn}$ as,
\bea
Q^{\mu\nu} &=&  \frac{{\bar n}^\mu {\bar n}^\nu}{\bar n^2} .
\eea
We, now, construct the third  projection tensor $R^{\mn}$,  with a constraint such that the sum of $R^{\mn}$,
$B^{\mn}$ and $Q^{\mn}$ gives the vacuum projection operator $V^{\mn}$ as
\be
R^{\mn} =V^{\mn} - B^{\mn} - Q^{\mn} =A^{\mn}-Q^{\mn}=g_{\perp}^{\mn}-\frac{P^{\mu}_{\perp}P^{\nu}_{\perp}}{P_{\perp}^2}. \label{sum_tb}
\ee
It can be checked easily that all the projection tensors satisfy the following properties,
\begin{subequations}
\begin{align}
P_\mu Z^{\mn} &=0, \\
Z^{\mu\lambda}Z_{\lambda}^{\nu} &= Z^{\mn},\\
Z^{\mn}Z_{\mn} &= 1 .
\end{align}
\end{subequations}
where $Z=B,R,Q$.
The three projection tensors are orthogonal to each other:
\begin{subequations}
\begin{align}
Z^{\mn} Y_{\mn} &= 0, \\
\end{align}
\end{subequations}
where $Z\ne Y$ and $Y=B,R,Q$.

Now we construct the fourth tensor as
\bea
N^{\mn}&=& \frac{\bar u^{\mu}\bar n^{\nu}+\bar u^{\nu}\bar n^{\mu}}{\sqrt{\bar u^2}\sqrt{\bar n^2}},
\eea
which satisfies the following properties
\bea
N^{\mu\rho}N_{\rho \nu}= B^{\mu}_{\nu}+Q^{\mu}_{\nu},
\eea
\bea
B^{\mu\rho}N_{\rho\nu}+N^{\mu\rho}B_{\rho\nu}=N^{\mu}_{\nu}\,\,,\,\,
Q^{\mu\rho}N_{\rho\nu}+N^{\mu\rho}Q_{\rho\nu}=N^{\mu}_{\nu},
\eea
\bea
R^{\mu\rho}N_{\rho\nu}=N^{\mu\rho}R_{\rho\nu}=0.
\eea
Now, one can write a general covariant structure of gauge boson self-energy
as 
\bea
\Pi^{\mn} = b B^{\mn} + c R^{\mn} + d Q^{\mn}+ a N^{\mn}, \label{gen_tb}
\eea
where $b$, $c$, $d$ and $a$ are four Lorentz-invariant form factors associated with
the four basis tensors. Note that Eq.~\eqref{gen_tb} can also be expressed as
\bea
\Pi^{\mn} = b B^{\mn} + c A^{\mn} + (d-c) Q^{\mn}+ a N^{\mn}
\eea
This particular decomposition of the self-energy in terms of four tensor basis is exactly same that has been used in Ref.~\cite{Romatschke:2003ms,Ghosh:2019fet} which, however
were then applied for different perspectives.

The $(00)$ components of the constituent tensors are given by
\begin{subequations}
\begin{align}
B_{00} &= \bar{u}^2,\\
R_{00} &= 0,\\
Q_{00} &= 0,\\
 N_{00}&=0,\\
 \Pi_{00} &= b B_{00} = b\bar{u}^2 .  
\label{debye_mass}
\end{align}
\end{subequations}
Using these information, we obtain the form factors as
\begin{subequations}
\begin{align}
b &= B^{\mn}\Pi_{\mn}, \label{ff_b} \\
c &= R^{\mn}\Pi_{\mn}, \label{ff_c} \\
d &= Q^{\mn}\Pi_{\mn}, \label{ff_d} \\
a &= \frac{1}{2}N^{\mn}\Pi_{\mn} \label{ff_a}.
\end{align}
\end{subequations} 
In absence of the magnetic field by comparing with the known general form 
of finite temperature self-energy in Eq.~(\ref{gen_struc_t}), 
as
\bea
\Pi_TA_{\mn}+\Pi_LB_{\mn} = b_0B_{\mn}+c_0R_{\mn}+d_0Q_{\mn}+a_0 N_{\mn},
\eea
one can write
\begin{subequations}
\begin{align}
b_0 &= \Pi_{L},\\
c_0&= d_0= \Pi_{T},\\
a_0&=0
\end{align}
\end{subequations} 
where we used the fact that $R_{\mn}+Q_{\mn}=A_{\mn}$. 


\section{General form of gauge boson propagator in   a hot magnetized medium}
\label{gsprop}

In covariant gauge the inverse of the gauge boson propagator in vacuum  reads as
\bea
\left(\mathcal{D}^0\right)^{-1}_{\mn} = P^2g_{\mn} - \frac{\xi -1}{\xi}P_\mu P_\nu,
\label{inverse_vacuum_prop_1t}
\eea
where $\xi$ is the gauge parameter.  From Eq.~(\ref{sum_tb}) one can write
\bea
P_\mu P_\nu &=& P^2 \big [ g_{\mn} -(B_{\mn} +R_{\mn} +Q_{\mn})\big ].
\eea
and using in Eq.~(\ref{inverse_vacuum_prop_1t}), we get
\bea
\left(\mathcal{D}^0\right)^{-1}_{\mn} = \frac{P^2}{\xi}g_{\mn} + P^2\frac{\xi -1}{\xi}\left(B_{\mn} +R_{\mn} +Q_{\mn}\right).
\label{inverse_vacuum_prop_2t}
\eea
The inverse of the general gauge boson propagator following  
 Dyson-Schwinger equation reads as
\bea
\mathcal{D}^{-1}_{\mn}= \left(\mathcal{D}^0\right)^{-1}_{\mn}  - \Pi_{\mn}.
\eea
From Eqs.~(\ref{inverse_vacuum_prop_2t}) and (\ref{gen_tb}) we can now readily get
\bea
\mathcal{D}^{-1} _{\mn}= \frac{P^2}{\xi}g_{\mn} +
\left(P_m^2 - b\right)B_{\mn} + \left(P_m^2 - c\right)R_{\mn} + 
\left(P_m^2 - d\right)Q_{\mn}-a N_{\mn}, 
\label{inverse_prop}
\eea
where 
\bea
P_m^2 = P^2\frac{\xi -1}{\xi}.
\eea
The inverse of Eq.~(\ref{inverse_prop}) can be written as
\bea
\mathcal{D}_{\mr} = \alpha P_\mu P_\rho + \beta B_{\mr} + \gamma R_{\mr} + \delta Q_{\mr}+ \sigma N_{\mr}. 
\eea
along with 
\bea
g_\mu^\nu= \mathcal{D}_{\mr} \left(\mathcal{D}^{-1}\right)^{\rn}  
&=& \alpha\frac{P^2}{\xi} P_\mu P^\nu  + \left[\frac{\beta P^2}{\xi}+\beta(P_m^2-b)-\sigma a\right]B_\mu^\nu \nn\\
&&  + \left[\frac{\delta P^2}{\xi}+\delta(P_m^2-d)-\sigma a\right]Q_\mu^\nu+\left[\frac{\gamma P^2}{\xi}+\gamma(P_m^2-c)\right]R_\mu^\nu\nn\\
&&+\left[-\beta a+\sigma (P_m^2-d)+\frac{\sigma P^2}{\xi}\right]\frac{\bar u_\mu \bar n^\nu}{\sqrt{\bar u^2}\sqrt{\bar n^2}}+\left[-\delta a +\sigma(P_m^2-b)+\frac{\sigma P^2}{\xi}\right]\nn\\
&&\times\frac{\bar n_\mu \bar u^\nu}{\sqrt{\bar u^2}\sqrt{\bar n^2}}.
\eea
Now using the explicit forms of $B_\mu^\nu , R_\mu^\nu $, $Q_\mu^\nu$ and $N_\mu^\nu$ and 
equating different coefficients from both sides yield the following conditions:
\bea
\alpha&=&\frac{\xi}{P^4},\nn\\
\frac{\beta P^2}{\xi}+\beta(P_m^2-b)-\sigma a&=&1,\nn\\
\frac{\delta P^2}{\xi}+\delta(P_m^2-d)-\sigma a&=&1,\nn\\
\frac{\gamma P^2}{\xi}+\gamma(P_m^2-c)&=&1,\nn\\
-\beta a+\sigma (P_m^2-d)+\frac{\sigma P^2}{\xi}&=&0,\nn\\
-\delta a +\sigma(P_m^2-b)+\frac{\sigma P^2}{\xi}&=&0.
\eea
Solving this we get
\bea
\alpha&=&\frac{\xi}{P^4},\nn\\
\beta&=& \frac{P^2-d}{(P^2-b)(P^2-d)-a^2},\nn\\
\gamma&=&\frac{1}{P^2-c},\nn\\
\delta&=&\frac{P^2-b}{(P^2-b)(P^2-d)-a^2},\nn\\
\sigma&=&\frac{a}{(P^2-b)(P^2-d)-a^2}.
\eea
Now the general covariant structure of the gauge boson propagator in covariant gauge can finally be obtained as 
\bea
\mathcal{D}_{\mn} &=&\frac{\xi P_{\mu}P_{\nu}}{P^4}+\frac{(P^2-d) B_{\mn}}{(P^2-b)(P^2-d)-a^2}+\frac{R_{\mn}}{P^2-c}+\frac{(P^2-b) Q_{\mn}}{(P^2-b)(P^2-d)-a^2}\nn\\
&&+\frac{a N_{\mn}}{(P^2-b)(P^2-d)-a^2}.
\label{gauge_prop}
\eea
We recall that the breaking of boost invariance due to finite temperature leads to two modes 
(degenerate transverse mode and plasmino). Now, the breaking of the
rotational invariance in presence of magnetic field lifts the degeneracy of the transverse modes which introduces an
 additional mode in the hot medium. These three dispersive modes of gauge boson can be seen from the poles of Eq.~\eqref{gauge_prop}. 
 The poles $(P^2-b)(P^2-d)-a^2=0$, lead to two dispersive modes. We call one mode $n^+$ with energy $\omega_{n^+}$ and the other one $n^-$ with energy $\omega_{n^-}$. The pole $P^2-c=0$
leads to the third dispersive mode $c$ with energy $\omega_c$. We will discuss about these dispersive modes in details later for both strong and weak field approximation.

When we turn off the magnetic field, the general structure of the propagator in a non-magnetized 
thermal bath can be obtained by putting $b_0=\Pi_L,\ c_0=d_0=\Pi_T$ and $a_0=0$ as
\bea
\mathcal{D}_{\mr} 
&=& \frac{\xi P_\mu P_\rho}{P^4} + \frac{B_{\mr}}{P^2-\Pi_L}  + \frac{A_{\mr}}{P^2-\Pi_T}\hspace{2cm}  
\eea
which agrees with the known result~\cite{Das;1997,  Lebellac;1996, Kapusta:1989tk,Andersen:2003zk}.

\section{Form Factors}
\label{ff}
Before computing the various form factors
associated with the general structure we note the following  points:

\begin{enumerate}
\item
The magnetic field  generated during the non-central HIC is time dependent but is   believed to 
decrease rapidly with time~\cite{Bzdak:2012fr,McLerran:2013hla}. It would  be extremely complicated to work with a time dependent 
magnetic field. Instead  we work  by considering a constant  background magnetic field  along with some limiting conditions 
so that the effect of magnetic field can be  incorporated analytically. We note here that incorporation of magnetic field to  the heat bath  introduces 
another scale in the system. Beside the fermion mass $m_f$ and the  temperature $T$,  the additional scale
is the strength of  magnetic field $B$. 
Below we would discuss the different  domains of scales:

{\sf   a)  Strong Field Approximation:} At the time of the collision, the value of the magnetic field $B$ 
is estimated upto the order of $|eB| \sim 15 m_\pi^2$ (where $e$ is the electronic charge, $m_\pi$ is the mass of a pion),  which  is very high compared 
to the temperature $T$  and $m_f$  in 
the LHC at CERN~\cite{Skokov:2009qp}. Also in the dense sector, neutron stars (NS), or more specifically magnetars are known to 
possess strong enough magnetic field~\cite{Duncan:1992hi,Chakrabarty:1997ef,Bandyopadhyay:1997kh}. The effect of this strong enough 
magnetic field can be  incorporated via a simplified Lowest Landau Level (LLL) approximation in which fermions are basically confined within 
the LLL. In the sec.~\ref{sfa} we will work on strong  field  approximation with a scale hierarchy, $m_f <  T < \sqrt{|eB|}$,  where 
 the loop momentum $K\sim T$ within HTL approximation.

{\sf  b) Weak Field Approximation:} Furthermore, it is believed that  the magnetic field generated in heavy-ion collisions decreases
rapidly with time.  This  provides us a simplified situation  where one can work in weak field approximation with a scale hierarchy, 
$\sqrt{|eB|} < m_f < T $ which will be discussed in details in subsection~\ref{wfa}.

\begin{figure}[tbh]
\vspace*{-0.1in}
\begin{center}
\includegraphics[scale=0.55]{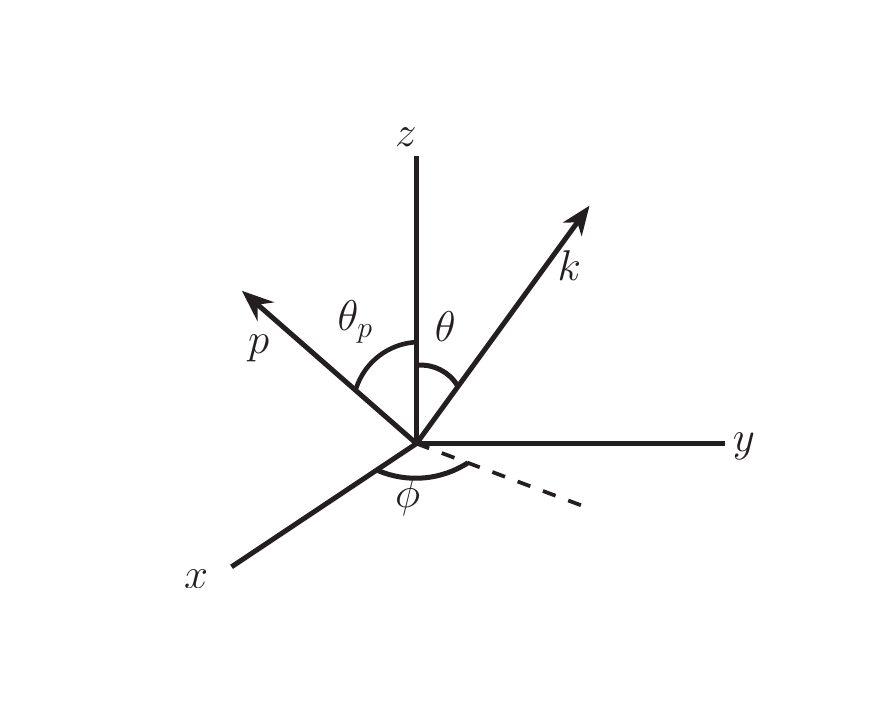}
\end{center}
\vspace*{-0.8in}
\caption{Choice of reference frame for computing the various form factors associated with the general structure of gauge boson 
2-point functions. The magnetic field is along  $z$-direction.}
\label{ref_frame}
\end{figure}

\item We would consider  $m_f=5 $ MeV for two light quark flavors $u$ and $d$. 
 
 \item We choose a frame of reference as shown in Fig.~\ref{ref_frame} in which one considers the external momentum of the vector 
 boson in $xz$ plane\footnote{However, one can consider a general frame of reference $P_{\mu}=(p_0,p_1,p_2,p_3)$
 and  the result would be independent of the choice of reference frame. Because 
 $p_{\perp}$ and $p_3$ are not in same footing due to the anisotropy caused by the external magnetic field along $z$ direction.
 But there is no distinction between $p_1$ and $p_2$. So, for simplicity  of the calculation,  we made a particular choice for 
 the reference frame here.}  with $0 <\theta_p< \pi/2$. So one can write 
\bea
P^\mu=(p_0,p\sin{\theta_p},0,p\cos{\theta_p}), \label{xy}
\eea
and then loop momenta 
\be
K^\mu=(k_0,k \sin{\theta} \cos{\phi},k \sin{\theta} \sin{\phi},k \cos{\theta}).
\ee
\end{enumerate}
\subsection{Gauge boson  in strongly magnetized medium}
\label{sfa}

\subsubsection{One-loop gluon self-energy}
\begin{figure}[tbh]
\begin{center}
\includegraphics[scale=0.55]{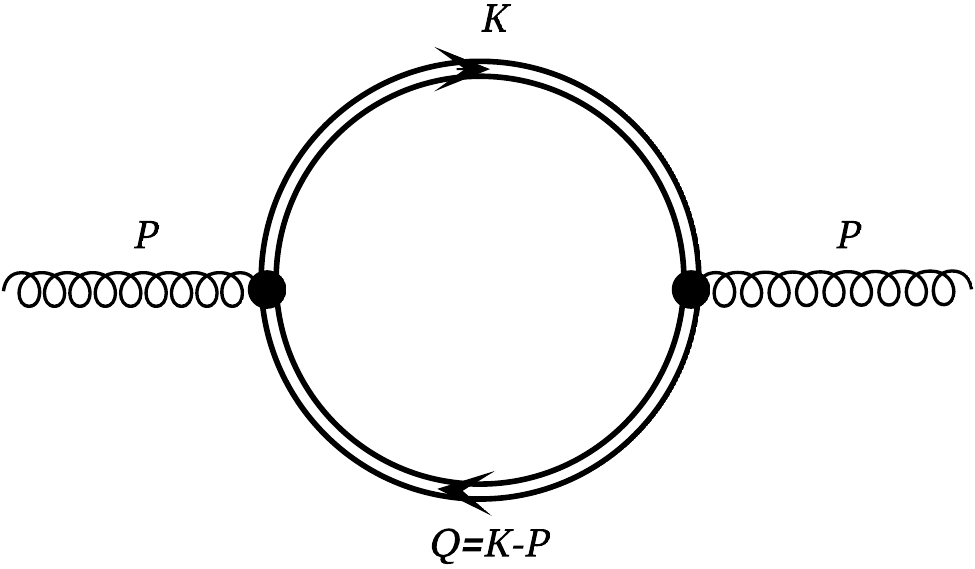}
\end{center}
\caption{Gluon polarization tensor in the limit of strong field approximation.}
\label{sfa_self_energy}
\end{figure}

When the external magnetic field is very strong, $  eB\rightarrow \infty$, 
it pushes all the Landau levels ($n\ge 1$) to infinity
compared to the Lowest Landau Level (LLL) with $n=0$. For 
LLL approximation in the strong field limit the 
fermion propagator reduces to a simplified form as
\bea
iS^s_{m}(K)=ie^{-{k_\perp^2}/{|q_fB|}}~~\frac{\slashed{K}_\sp+m_f}{
K_\sp^2-m_f^2}(1-i\gamma_1\gamma_2),
\label{prop_sfa}
\eea
where $K$ is the fermionic four momentum and we have used the properties of generalized Laguerre polynomial, $L_n\equiv L_n^0$ and $L_{-1}^\alpha = 0$. In strong field approximation or in LLL, $eB \gg k_\perp^2$, an effective dimensional reduction from
$(3+1)$ to $(1+1)$ takes place.

Now in the strong field limit the self-energy\,(Fig.~\ref{sfa_self_energy}) can be computed as
\bea
\Pi_{\mu\nu}^s(P) &=& \sum_f\frac{i  
  g^2}{2}\int\frac{d^4K}{(2\pi)^4}\textsf{Tr}\left[\gamma_\mu S^s_{m}(K)\gamma_\nu 
S^s_{m}(Q)\right]\nn\\
&=& \sum_f\frac{ig^2}{2} \int\frac{d^2k_\perp}{(2\pi)^2} 
\exp\left(\frac{-k_\perp^2-q_\perp^2}{|q_fB|}\right)\nn\\
&&\times \int\frac{d^2K_\sp}{(2\pi)^2} \textsf{Tr} \left[\gamma_\mu \frac{\slashed{K}_\sp+m_f}
{K_\sp^2-m_f^2}(1-i\gamma_1\gamma_2)\gamma_\nu  
\frac{\slashed{Q}_\sp+m_f}{Q_\sp^2-m_f^2}(1-i\gamma_1\gamma_2)\right
],
\eea
where `\textit{s}' indicates that the quantities are to be calculated in the strong field approximation and $\textsf{Tr}$ represents only the Dirac trace. We have suppressed the 
color indices for convenience. Now one can notice that the longitudinal and transverse parts are completely separated and the Gaussian integration over the transverse momenta can be done trivially, which leads to
\bea
\Pi_{\mu\nu}^s(P)&=& \sum_f i 
  ~e^{{-p_\perp^2}/{2|q_fB|}}~\frac{g^2 |q_fB|}{2\pi}\int\frac{d^2K_\sp}{(2\pi)^2} 
\frac{{\cal S}_{\mu\nu}^s}{(K_\sp^2-m_f^2)(Q_\sp^2-m_f^2)} \nn\\
&=& ~-\sum_fe^{{-p_\perp^2}/{2|q_fB|}}~\frac{g^2 |q_fB|}{2\pi}~T\sum\limits_{k_0}\int\frac{dk_3}{2\pi} 
\frac{{\cal S}_{\mu\nu}^s}{(K_\shortparallel^2-m_f^2)(Q_\shortparallel^2-m_f^2)}, 
\label{pol_vacuum}
\eea
with the tensor structure ${\cal S}_{\mu\nu}^s$ that originates from the Dirac trace is 
\bea
{\cal S}_{\mu\nu}^s = K_\mu^\shortparallel Q_\nu^\shortparallel + Q_\mu^\shortparallel K_\nu^\shortparallel 
- g_{\mu\nu}^\shortparallel \left((K\cdot Q)_\shortparallel -m_f^2\right),
\eea
where the Lorentz indices  $\mu$ and $\nu$ are restricted to longitudinal values 
because 
of dimensional reduction to (1+1) dimension and forbids to take any transverse values.
Now we use Eq.~(\ref{p_sp}) and Eq.~(\ref{eta_sp}) to rewrite $S_{\mn}$ as
\bea
{\cal S}_{\mu\nu}^s &=& (k_0 u_\mu - k_3 n_\mu ) (q_0 u_\nu - q_3 n_\nu )  + (q_0 u_\mu - q_3 n_\mu )  (k_0 u_\nu - k_3 n_\nu ) \nn\\
&&- (u_\mu u_\nu - n_\mu n_\nu)\left((k\cdot q)_\shortparallel -m_f^2\right) \nn\\
&=& u_\mu u_\nu  \left( k_0 q_0 + k_3 q_3 +m_f^2\right) + n_\mu n_\nu  \left( k_0 q_0 + k_3 q_3 -m_f^2\right)\nn\\
&& - \left( u_\mu n_\nu + n_\mu u_\nu \right) \left( k_0 q_3 + k_3 q_0 \right).
\label{se_sfa_gen}
\eea
\subsubsection{Form factors and Debye mass}
First we evaluate the form factors in Eqs.~(\ref{ff_b}),~(\ref{ff_c}) ,~(\ref{ff_d}) and ~(\ref{ff_a}) 
in strong field approximation as 
\begin{subequations}
\begin{align}
c &= R^{\mn}(\Pi_{\mn}^{g}+\Pi_{\mn}^s) =c_{YM}+c_s=\frac{C_Ag^2T^2}{3}\frac{1}{2}\left[\frac{p_0^2}{p^2}-\frac{P^2}{p^2}\mathcal{T}_P(p_0,p)\right] \,\mbox{where\,\,} c_s=0, \label{coeff_1}\\
b &= B^{\mn}(\Pi_{\mn}^{g}+\Pi_{\mn}^s)
= b_{YM}+\frac{u^\mu u^\nu}{\bar{u}^2} \Pi_{\mn}^s=b_{YM}+b_s\nn\\
&=\frac{C_Ag^2T^2}{3\bar u^2}\left[1-\mathcal{T}_P(p_0,p)\right]-\sum_fe^{{-p_\perp^2}/{2|q_fB|}}~~\frac{g^2 |q_fB|}{2\pi 
\bar{u}^2}~T\sum\limits_{k_0}\int\frac{dk_3}{2\pi} 
\frac{k_0 q_0 + k_3q_3 
+m_f^2}{(K_\shortparallel^2-m_f^2)(Q_\shortparallel^2-m_f^2)}, \label{coeff_2}\\
d &= d_{YM}+Q^{\mn}\Pi_{\mn}^s
= d_{YM}+d_s\\
&=\frac{C_Ag^2T^2}{3}\frac{1}{2}\left[\frac{p_0^2}{p^2}-\frac{P^2}{p^2}\mathcal{T}_P(p_0,p)\right]+\sum_f e^{{-p_\perp^2}/{2  |q_fB|}}~\frac{g^2|q_f B|}{2\pi}\frac{p_{\perp}^2}{p^2}~ T \sum_{k_0}\int\frac{dk_3}{2\pi} \frac{k_0 q_0+k_3q_3-m_f^2}{(K_\sp^2-m_f^2)(Q_\sp^2-m_f^2)},\label{coeff_3}\\
a &=\frac{1}{2}N^{\mn}(\Pi_{\mn}^{g}+\Pi^s_{\mn})=\frac{1}{2}N^{\mn}\Pi^s_{\mn}=a_s,\,\mbox{where\,\,}a_{YM}=0,\label{coeff_4}
\end{align}
\end{subequations}
where $\Pi_{\mn}^{g}$ is the Yang-Mills(YM) contribution from ghost and gluon loop which remains unaffected in presence of magnetic field and can be written as
\bea
\Pi_{\mn}^{g}(P)=-\frac{N_cg^2T^2}{3} \int \frac{d \Omega}{2 \pi}\left(\frac{p_{0} \hat{K}_{\mu} \hat{K}_{\nu}}{\hat{K} \cdot P}-g_{\mu 0} g_{\nu 0}\right).\label{Pi_mn_g}
\eea

Now, combining Eq.~(\ref{coeff_2}) and the  Hard Thermal Loop (HTL) 
approximation~\cite{pisarski:1990} one can have 
\bea
b_s
&\approx& -\sum_fe^{{-p_\perp^2}/{2|q_fB|}}~\frac{g^2|q_fB|}{2\pi 
	\bar{u}^2}~T\sum\limits_{k_0}\int\frac{dk_3}{2\pi} \left[\frac{1}{(K_\shortparallel^2-m_f^2)}+\frac{2\left(k_3^2+m_f^2\right)}
	{(K_\shortparallel^2-m_f^2)(Q_\shortparallel^2-m_f^2)}\right]\nn\\
&=& \sum_fe^{{-p_\perp^2}/{2|q_fB|}}~\frac{g^2|q_fB|}{2\pi 
	\bar{u}^2}~\int\frac{dk_3}{2\pi} \Bigg[-\frac{n_F(E_{k_3})}{E_{k_3}}\nn\\
&&\hspace{2cm}+\left\{\frac{n_F(E_{k_3})}{E_{k_3}}+\frac{p_3 k_3}{E_{k_3}}\frac{\partial n_F(E_{k_3})}{\partial k_3}\left(\frac{p_3 k_3/E_{k_3}}{p_0^2-p_3^2(k_3/E_{k_3})^2}\right)\right\}\Bigg]\nn\\
&=& \sum_fe^{{-p_\perp^2}/{2|q_fB|}}~\frac{g^2|q_fB|}{2\pi 
	\bar{u}^2}\int\frac{dk_3}{2\pi}\,\frac{p_3 k_3}{E_{k_3}} \frac{\partial n_F(E_{k_3})}{\partial E_{k_3}}\left(\frac{p_3 k_3/E_{k_3}}{p_0^2-p_3^2(k_3/E_{k_3})^2}\right) . \label{b_m_sf}
\eea
Using Eq.~(\ref{coeff_2}),(\ref{b_m_sf}) in  Eq.~(\ref{debye_mass}) one also can directly calculate the Debye screening mass in QCD as
\bea
(m_D^2)_s &=& \left.{\bar u}^2 b \right |_{p_0=0,\ \bm{p} \rightarrow 0}=m_D^2+\sum_f (\delta m_{D,f}^2)_s
= m_D^2 - \sum_f\frac{g^2|q_fB|}{2\pi }~\int\frac{dk_3}{2\pi} \frac{\partial n_F(E_{k_3})}{\partial E_{k_3}} \nn\\
&=& \frac{g^2N_c T^2}{3}+ \sum_f\frac{g^2|q_fB|}{2\pi  T}~\int_{-\infty}^\infty\frac{dk_3}{2\pi}~ n_F(E_{k_3})\left(1-n_F(E_{k_3})\right ).
 \label{dby_m_sf}
\eea 
which reduces to the expression of QED Debye mass calculated in Refs.~\cite{Alexandre:2000jc,Bandyopadhyay:2016fyd} without QCD factors
where three distinct scales ($m_f^2$, $T^2$ and $eB$) were clearly 
evident for massive quarks.

Now using Eq.~(\ref{dby_m_sf}) in Eq.~(\ref{b_m_sf}) along with $E_{k_3}\sim k_3$, the form factor $b$ can be expressed  in terms of $m_D$ as
\begin{equation}
 b =\frac{C_Ag^2T^2}{3\bar u^2}\left[1-\mathcal{T}_P(p_0,p)\right]-\sum_f e^{{-p_\perp^2}/{2  |q_fB|}}~\left(\frac{\delta m_{D,f}}{\bar u}\right)^2\frac{p_3^2}{p_0^2-p_3^2}\label{b_sf} .\
\end{equation}

The form factor $d$  then becomes
\begin{equation}
 d \approx \frac{C_Ag^2T^2}{3}\frac{1}{2}\left[\frac{p_0^2}{p^2}-\frac{P^2}{p^2}\mathcal{T}_P(p_0,p)\right]+\sum_f e^{{-p_\perp^2}/{2 |q_fB|}}~\left(\frac{\delta m_{D,f}}{\bar u}\right)^2\frac{p_3^2}{p_0^2-p_3^2} . \label{d_m_sf}
\end{equation}
where the expression for $(\Pi_\mu^\mu)^s$  is given  in Eq.~(\ref{pimm_result})  in Appendix~\ref{sfa_app}.

The form factor $d_s$ can be calculated as
\bea
d_s&=&Q^{\mn} \Pi_{\mn}^s,\nn\\
&\approx&-\sum_f i~e^{{-p_\perp^2}/{2  |q_fB|}}~\frac{g^2|q_f B|}{2\pi}\frac{p_{\perp}^2}{p^2}\int\frac{d^2K_\sp}{(2\pi)^2} \bigg[\frac{\big(k_0^2+k_3^2-m_f^2\big)}{(K_\sp^2-m_f^2)(Q_\sp^2-m_f^2)}\bigg],nn\\
&\approx&\sum_f e^{{-p_\perp^2}/{2 |q_fB|}}~\delta m_{D,f}^2~\frac{p_{\perp}^2}{p^2}\frac{p_3^2}{p_0^2-p_3^2} \label{d_contract}
\eea
for $k_3 \sim E_{k_3}$.
Now using \eqref{d_contract} in \eqref{coeff_3}, the form factor $d$ can be written as
\bea
d \approx \frac{C_Ag^2T^2}{3}\frac{1}{2}\left[\frac{p_0^2}{p^2}-\frac{P^2}{p^2}\mathcal{T}_P(p_0,p)\right]+\sum_f e^{{-p_\perp^2}/{2 |q_fB|}}~\delta m_{D,f}^2~\frac{p_{\perp}^2}{p^2}\frac{p_3^2}{p_0^2-p_3^2} , \label{d_m_sf}
\eea
where $p_3=p\cos\theta_p$ and $p_\perp=p\sin\theta_p$ as given in Eq.~\eqref{xy}.

Also
\bea
2a=N^{\mn} \Pi_{\mn}^s&=&\sum_fi~e^{{-p_\perp^2}/{2  |q_fB|}}~\frac{g^2|q_f B|}{2\pi \sqrt{\bar u^2}\sqrt{\bar n^2}}\int\frac{d^2K_\sp}{(2\pi)^2} \bigg[\frac{-2\frac{\bar u \cdot n}{\bar u^2}\big(k_0^2+k_3^2+m_f^2\big)+4k_0k_3}{(K_\sp^2-m_f^2)(Q_\sp^2-m_f^2)}\bigg]\nn\\
&=&\sum_f e^{{-p_\perp^2}/{2  |q_fB|}}~\frac{g^2|q_f B|}{2\pi \sqrt{\bar u^2}\sqrt{\bar n^2}}\int\frac{dk_3}{2\pi}\bigg[-2\frac{\bar u \cdot n}{\bar u^2} \frac{\partial n_F(E_{k_3})}{\partial E_{k_3}}\frac{p_3^2 k_3^2/E_{k_3}^2}{\big(p_0^2-p_3^2 k_3^2/E_{k_3}^2\big)}\nn\\
&&+\frac{2\partial n_F(E_{k_3})}{\partial E_{k_3}}\frac{p_0p_3 k_3^2/E_{k_3}^2}{\big(p_0^2-p_3^2 k_3^2/E_{k_3}^2\big)}\bigg] \nn\\
&\approx& \sum_f 2 ~e^{{-p_\perp^2}/{2  |q_fB|}} \frac{\sqrt{\bar n^2}}{\sqrt{\bar u^2}}\delta m_{D,f}^2~ \frac{p_0p_3}{p_0^2-p_3^2},\label{a_m_sf}
\eea
where   $\bar n^2=-p_{\perp}^2/p^2=-\sin^2{\theta_p}$ and $\bar u^2=-p^2/P^2$.

Also in the strong field approximation, $ |eB| > T^2 > m_f^2$,  one can neglect the 
quark mass $m_f$, to get an  
analytic expression of Debye mass as 
\bea
(m_D^2)_s&=&\frac{g^2N_c T^2}{3}+ \sum_f\frac{g^2|q_fB|}{2\pi  T}~\int_{-\infty}^\infty\frac{dk_3}{2\pi}~ n_F(k_3)\left(1-n_F(k_3)\right )\nn\\
&=&\frac{g^2N_c T^2}{3}+\sum_f \frac{g^2 |q_fB|}{4\pi^2}\nn\\
&=&m_D^2+\sum_f (\delta m_{D,f}^2)_s\nn\\
&=&m_D^2+(\delta m_D^2)_s,\label{debye2_mass}
\eea
which agrees with that obtained in Ref.~\cite{Bandyopadhyay:2016fyd}.

\subsubsection{Dispersion}
As discussed after Eq.~(\ref{gauge_prop}),  the dispersion relations  for gluon in strong field approximation with  LLL 
read as
\begin{subequations}
\begin{align}
P^2-c &= 0,\\
(P^2-b)(P^2-d)-a^2&=(P^2-\omega_n^+)(P^2-\omega_n^-)=0,
\end{align}
\end{subequations}
with
\begin{subequations}
\begin{align}
 \omega_{n^+}&=\frac{b+d+\sqrt{\(b-d\)^2+4a^2}}{2},\\
\omega_{n^-}&=\frac{b+d-\sqrt{\(b-d\)^2+4a^2}}{2},
\end{align}
\end{subequations}
where  the form factors are given, respectively,  in Eqs.~(\ref{coeff_1}),~(\ref{b_sf}),~(\ref{d_m_sf}) and~(\ref{a_m_sf}).  

\begin{center}
 \begin{figure}
 \begin{center}
  \includegraphics[scale=0.32]{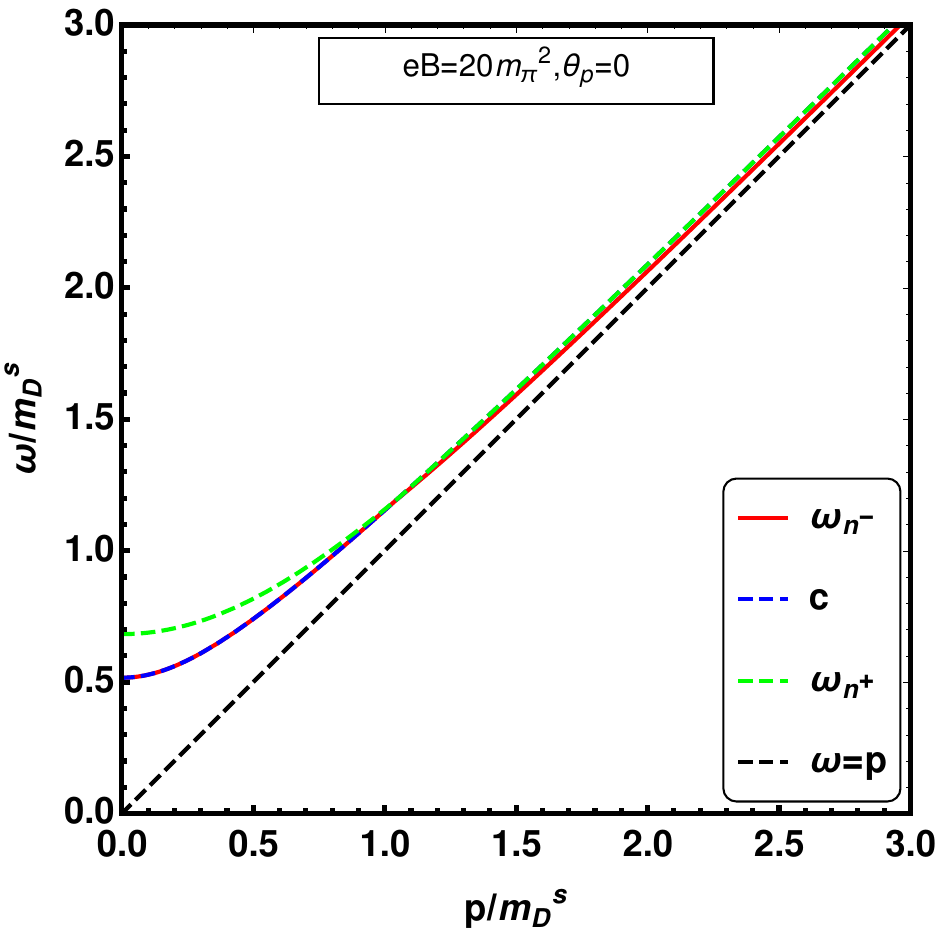}
  \includegraphics[scale=0.32]{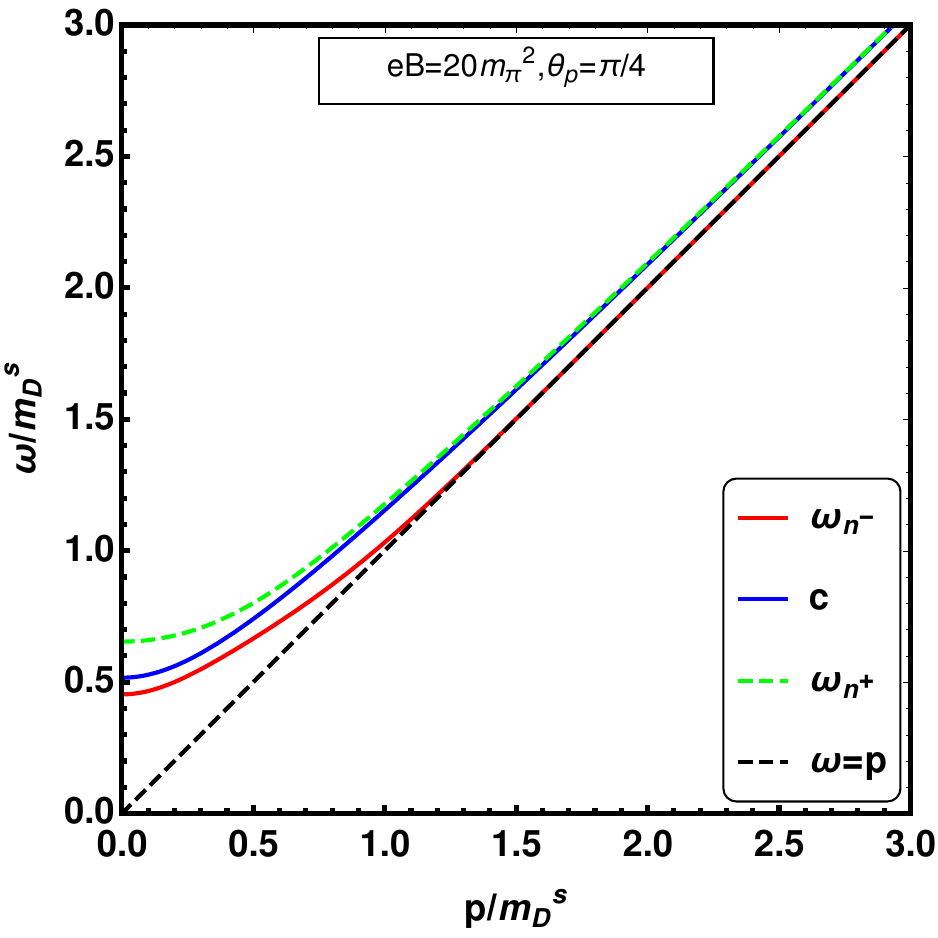}
  \includegraphics[scale=0.32]{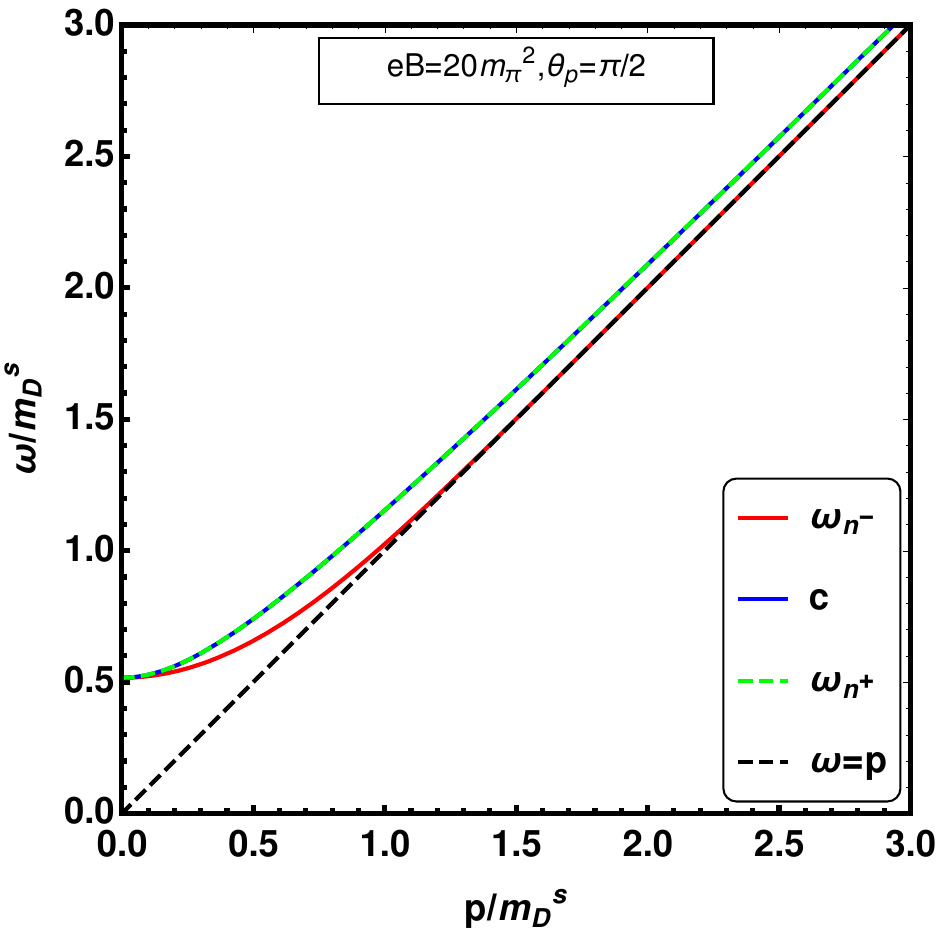}
  \caption{The plot of dispersion of the three modes ($n^-$, $c$ and $n^+$ modes) of a gauge boson in strong field  approximation 
  for propagation angles $\theta_p=0,\,\, \pi/4$ and $\pi/2$ at $eB=20m_{\pi}^2$ and $T=0.2$ GeV. $\omega=p$ represents the light cone.}
  \label{b_d_mode}
  \end{center}
 \end{figure}
\end{center}

The solutions of above three dispersion relations are named as $c$-mode, $n^+$-mode and $n^-$-mode with energies 
$\omega_c$, $\omega_{n^+}$ and $\omega_{n^-}$, respectively.  The dispersion plot for the three modes of gluon in strong field approximation is shown in Fig.~\ref{b_d_mode} for $|eB|=20m_{\pi}^2$, $T=0.2 $ GeV and for 
three propagation angles $\theta_p=0,\,\, \pi/4$ and $\pi/2$. We have used both magnetic field and temperature dependent coupling constant~\cite{Bandyopadhyay:2017cle} for the purpose. As found $c_s=0$ in Eq.~\eqref{coeff_1}  which implies that the $c$-mode  is unaffected
by the magnetic field and propagates like HTL transverse mode irrespective of the propagation angle as shown in Fig.~\ref{b_d_mode}.
The reason for which could be understood in the following way:
in strong field approximation there is an effective dimensional reduction from (3+1) 
to $(1+1)$ dimension in LLL.
Fermions at LLL can move only along the direction of external magnetic field. The electric field corresponding to the $c$ mode is always transverse to the external magnetic field irrespective of the propagation angle of gluon. Thus, the fermions are not affected by the gluon excitation~\cite{Hattori:2017xoo} and the quark loop contribution ($c_s$) becomes zero.

Now we note that at $\theta_p=0$ the form factor $a=0$  as it is proportional to $\sin\theta_p\cos\theta_p$.
In this case both  $n^-$ and $c$ modes are degenerate as the form factors  coincide with the HTL $\Pi_T$ without the quark loop contribution. This is because quark loop contribution
 in the form factor $d$ in Eq.~\eqref{d_m_sf} is proportional to $ \sin^2\theta_p\cos^2\theta_p$.
 This makes $n^-$ and $c$ mode to coincide with the HTL transverse dispersive mode. 
 This can be seen from the left panel of  Fig.~\ref{b_d_mode}. It could also be understood in the following way:
 when gluon propagates along the direction of external magnetic field, {\it{i.e}}, $\theta_p=0$, the two transverse modes become rotationally symmetric about the external magnetic field and become degenerate which is shown in the left panel of Fig.~\ref{b_d_mode}. The electric fields corresponding to the $n^-$ and $c$ modes are perpendicular to the external magnetic field. Thus two transverse electric fields can not excite the fermions whose movement are restricted to the direction of external magnetic field in LLL~\cite{Hattori:2017xoo}. This makes the quark loop contribution zero as noted earlier. In addition to the two transverse modes $n^-$ and $c$, there is also a longitudinal excitation $n^+$ at $\theta_p=0$. At any intermediate angle of propagation, {\it{e.g}}, $\theta_p=\pi/4$, the degeneracy of the transverse modes is lifted as shown in the middle panel of Fig.~\ref{b_d_mode}. Here both the transverse and longitudinal modes can excite the fermions as the corresponding electric fields are not orthogonal to the external magnetic field. As the propagation angle increases, the pole position corresponding to the $n^-$ mode shifts from transverse channel and approaches the longitudinal channel~\cite{Hattori:2017xoo}. At $\theta_p=\pi/2$, the
 form factor $a$ in Eq.~\eqref{a_m_sf} and the quark contribution of the form factor $d$ in Eq.~\eqref{d_m_sf} also vanish because of their $\theta_p$ dependence. Thus, the $n^-$ mode merges with HTL longitudinal mode whereas the $n^+$ mode merges with $c$ mode. This is reflected in the right panel of Fig.~\ref{b_d_mode}.

 \subsection{Gauge boson in weakly magnetized hot medium}
 \label{wfa}

\subsubsection{One-loop gluon self-energy}
\begin{center}
 \begin{figure}[tbh]
 \begin{center}
  \includegraphics[scale=0.5]{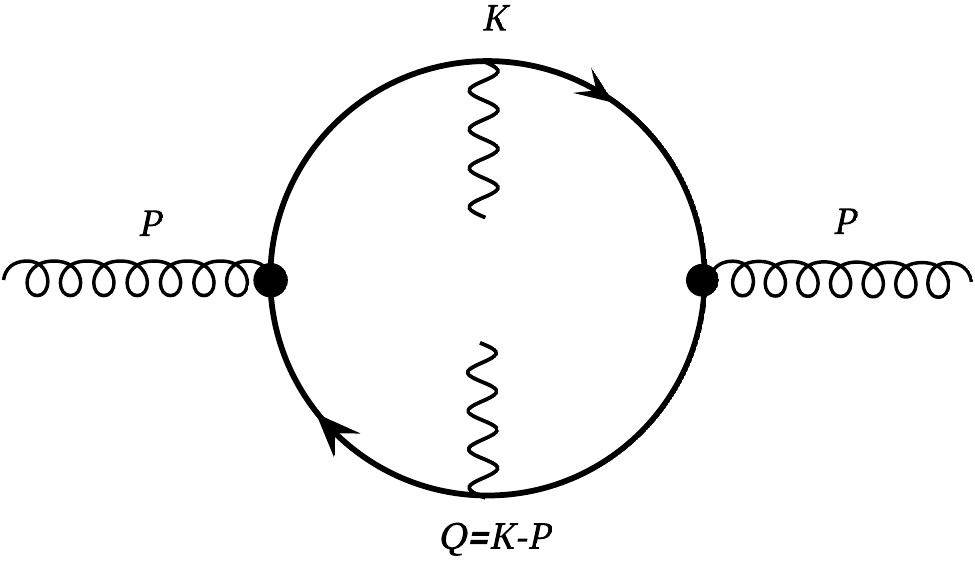}
  \caption{The order of $(eB)^2$ correction to the gluon polarization tensor $(\delta\Pi_{\mu\nu}^{a})$ in weak field approximation.}
  \label{wfa_se}
  \end{center}
 \end{figure}
\end{center}

The fermion propagator in a weak magnetic field, \textit{i.e.,}  $\sqrt{|eB|} <(K\sim  T)$ and $m_f$,  can be written up to $\mathcal{O}[(eB)^2]$ 
as
\bea
iS^w_m(K) &=&  i\frac{\slashed{K}+m_f}{K^2-m_f^2} +i ~(q_f B)\frac{\left(\gamma_5
\left\{(K\cdot n)\slashed{u}-(K\cdot u)\slashed{n}\right\}+i\gamma_1\gamma_2m_f\right)}{(K^2-m_f^2)^2} \nn\\
&+& i \ 2(q_fB)^2  \left[\frac{\left\{(K\cdot u)\slashed{u}-(K\cdot n)\slashed{n}\right\} 
-\slashed{K}}{(K^2-m_f^2)^3}-\frac{k_\perp^2(\slashed{K}+m_f)}{(K^2-m_f^2)^4}\right]  
+ \mathcal{O}\left[(e B)^3\right]\nn\\
&=& S_0 +S_1+ S_2+ \mathcal{O}[(e B)^3],
\label{prop_wfa}
\eea
where $S_0$ is the continuum free field propagator in absence of $B$ whereas $S_1$ 
and $S_2$ are, respectively,  $\mathcal{O}[(eB)]$ and  $\mathcal{O}[(eB)^2]$ 
correction terms in presence of $B$.  
The contribution to the gluon self-energy due to the quark loop can be written from the Feynman diagram~Fig.~\ref{wfa_se} as 
\bea
\Pi^{w,q}_{\mu\nu}(P)&=&\sum_f \frac{ig^2}{2}\int\frac{d^4K}{(2\pi)^4}\textsf{Tr}\left [\gamma_\mu S^w_m(K)\gamma_\nu S^w_m(Q)\right]. 
\eea

\begin{center}
 \begin{figure}[tbh]
 \begin{center}
  \includegraphics[scale=0.5]{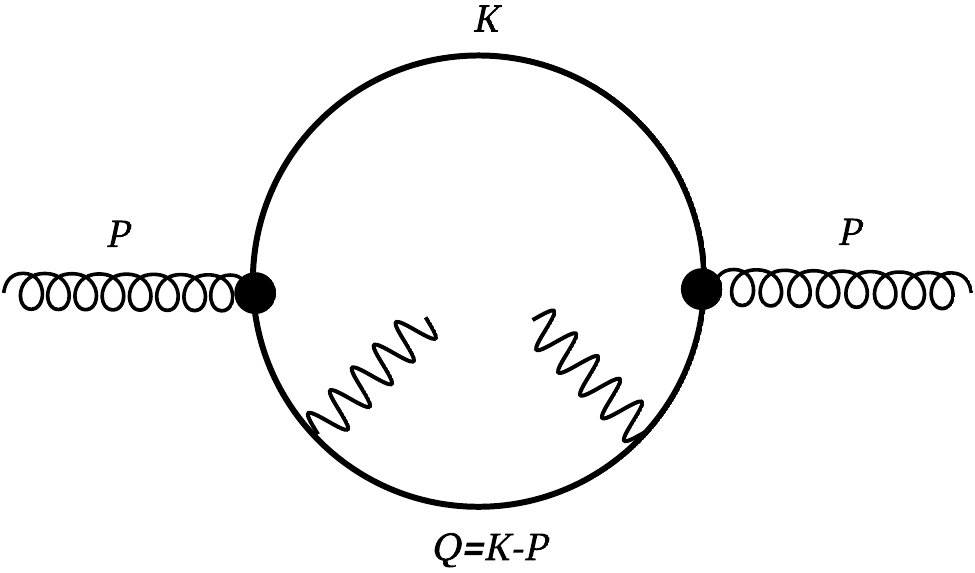} \hspace*{0.1in}
  \caption{The order of $(eB)^2$ correction to the gluon polarization tensor $(\delta\Pi_{\mu\nu}^{b})$
  in weak field approximation.}
  \label{wfa_vertex}
  \end{center}
 \end{figure}
\end{center}

We have suppressed the 
color indices here also for convenience. Using Eq.~(\ref{prop_wfa}) the self-energy in weak field approximation 
upto an $\mathcal{O}[(eB)^2]$ and also adding pure YM contribution, total gluon self-energy in weak field approximation can be decomposed as
\bea
\!\!\!\!\!\!\!\!\!\!\!
\Pi^w_{\mu\nu}(P)=  \Pi_{\mu\nu}^{g}(P) + \Pi_{\mu\nu}^{0}(P) \!+\delta\Pi_{\mu\nu}^{a}(P) +2\delta\Pi_{\mu\nu}^{b}(P) 
+\! \mathcal{O}[(eB)^3], 
\label{se_wfa}
\eea
where the first term  $\Pi_{\mn}^{g}$ is the YM contribution which is given in Eq.~\eqref{Pi_mn_g}.
The last three terms in Eq.~\eqref{se_wfa} appear from the expansion of quark loop contribution to the gluon self-energy. The term $\Pi_{\mn}^{0}$, containing two $S_0$ , is the 
leading order perturbative term in absence of $B$ whereas 
the remaining two terms are  $(eB)^2$ order corrections as shown in Fig.~\ref{wfa_se},\ref{wfa_vertex}. However, we note 
that $\mathcal{O}[(eB)]$ vanishes according to Furry's theorem since the expectation 
value of any odd number of electromagnetic currents must vanish due 
to the charge conjugation symmetry.  

Now the second and third terms in Eq.~(\ref{se_wfa}) can be written as
\bea
\Pi_{\mu\nu}^{0}(P)&=&\sum_f\frac{ig^2}{2}\int\frac{d^4K}{(2\pi)^4}\textsf{Tr}\left[\gamma_\mu S_0(K)\gamma_\nu S_0(Q)\right]\nn\\
&=&\sum_f\frac{ig^2}{2}\int\frac{d^4K}{(2\pi)^4}\left [8K_\mu K_\nu-4K^2g_{\mu\nu}\right]\frac{1}{{(K^2-m_f^2)(Q^2-m_f^2)}}, 
\label{se_pi_00} \\
 \delta\Pi_{\mu\nu}^{a}(P) &=& \sum_f\frac{ig^2}{2}\int\frac{d^4K}{(2\pi)^4}\textsf{Tr}\left [\gamma_\mu S_1(K)\gamma_\nu S_1(Q)\right ],\nn\\
&=& \sum_f\frac{ig^2}{2}(q_fB)^2\!\! \int\frac{d^4K}{(2\pi)^4} \frac{U_{\mu\nu}}{(K^2-m_f^2)^2(Q^2-m_f^2)^2}, \label{se_pi_11}
\eea
where in the numerator we have neglected the mass of the quark and the external momentum $P$ due to HTL approximation. The tensor structure of the  self-energy correction in weak field approximation comes out to be
\bea
U_{\mu\nu}  &=&  4(K \cdot u) (Q \cdot u)\left(2n_\mu n_\nu + g_{\mu\nu}\right)+4(K \cdot n) (Q \cdot n) \left(2u_\mu u_\nu - g_{\mu\nu}\right) \nn\\
&& - 4\left [(K \cdot u)(Q \cdot n)+(K \cdot n)(Q \cdot u)\right ] \left(u_\mu n_\nu + u_\nu n_\mu\right) +4m_f^2 g_{\mu\nu}  \nn \\
&& + 8m_f^2 \left (g_{1\mu}g_{1\nu} +g_{2\mu}g_{2\nu}\right ) . \label{U}
\eea

The third term in Eq.~(\ref{se_wfa}) can be written as
\bea
\!\!\!\! \!\!\!\! \delta \Pi_{\mu\nu}^{b}(P) &=& \sum_f\frac{ig^2}{2}\int\frac{d^4K}{(2\pi)^4}\textsf{Tr}\left [\gamma_\mu S_2(K)\gamma_\nu S_0(Q)\right ] \nn\\
&=&\sum_f ig^2(q_fB)^2\!\!\int\!\!\frac{d^4K}{(2\pi)^4}\!\!\left[\!\frac{X_{\mu\nu}}{(K^2-m_f^2)^3(Q^2-m_f^2)}\!-\!\frac{(K_\shortparallel^2-m_f^2) 
W_{\mu\nu}}{(K^2-m_f^2)^4(Q^2-m_f^2)}\!\right]
\eea
where
\begin{subequations}
\begin{align}
X_{\mu\nu} &= 4 \left[(K\cdot u) \left(u_\mu Q_\nu + u_\nu Q_\mu  \right) 
- (K\cdot n) \left(n_\mu Q_\nu + n_\nu Q_\mu\right) \right.\nn \\
&\left.  + \left\{ (K\cdot n)(Q\cdot n) - (K\cdot u)(Q\cdot u) +m_f^2 \right\} g_{\mu\nu}  \right], \label{X}\\
W_{\mu\nu} &= 4\left (K_\mu Q_\nu + Q_\mu K_\mu \right ) - 4\left (K\cdot Q-m_f^2 \right ) g_{\mu\nu}. \label{W}
\end{align}
\end{subequations}
\subsubsection{Computation of form factors and Debye mass of 
\texorpdfstring{$\mathcal{O}\[(eB)^0\]$}{[O(eB)0]} term}
In this subsection, we calculate the $\mathcal{O}\[(eB)^0\]$ terms in the form factors $b,c,d$ in the weak magnetic field limit 
 which are  denoted by $b_0,c_0,d_0$, respectively.  

The form factor $b_0$ in absence of magnetic field  can be written from Eq.~(\ref{debye_mass}) as
\bea
b_0(p_0,p)&=&\frac{1}{\bar{u}^2}\bigg[\Pi^{g}_{00}(P) +\Pi^{0}_{00}(P)\bigg] \label{b0}.
\eea
where  
\bea
\Pi^{0}_{00}(P)&=&\sum_f\frac{ig^2}{2}\int\frac{d^4K}{(2\pi)^4}\left [8k_0^2-4K^2\right]\frac{1}{{(K^2-m_f^2)(Q^2-m_f^2)}} .
\eea
 Using the hard thermal loop (HTL) approximation~\cite{Lebellac;1996}  and performing the frequency sum, one can write
\bea
\Pi^{0}_{00}(P) &=&-2g^2N_f\int\frac{k^2dk}{2\pi^2}~\frac{dn_F(k)}{dk}\int\frac{d\Omega}{4\pi}\left(1-\frac{p_0}{P\cdot\hat{K}}\right) , \label{b01}
\eea
for $ m_f=0$ . 

Now the  QCD Debye mass in the absence of the magnetic field can directly be obtained using  Eq.~\eqref{debye_mass}  as
\bea
m_D^2& = &\Pi^{0}_{00}\Big\vert_{p_0=0 \atop{\bf p} \rightarrow 0} = 
{\bar{u}^2} 
b_0\Big\vert_{p_0=0\atop {\bf p}\rightarrow  0}   = \frac{N_cg^2T^2}{3}-2g^2\int\frac{k^2dk}{2\pi^2}~\frac{dn_F(k)}{dk} =\frac{g^2T^2 }{3}\(N_c+\frac{N_f}{2}\) \label{debye_mass_qed}.\ 
\eea

Using Eq.~(\ref{debye_mass_qed}) in Eq.~(\ref{b01}), we get 
\bea
\Pi^{0}_{00}(P)
&=&\frac{N_fg^2T^2}{6}\int\frac{d\Omega}{4\pi}\(1-\frac{p_0}{p_0-\bm{p}\cdot\hat{\bm{k}}}\)
=\frac{N_fg^2T^2}{6}\(1-\frac{p_0}{2p}\log\frac{p_0+p}{p_0-p}\),
\label{Pi00_0}
\eea
where we use $p=\sqrt{p_1^2+p_3^2}$ as $p$ lies in $xz$ plane as shown Fig.~\ref{ref_frame}. The form factor in Eq.~(\ref{b0}) becomes
\bea
b_0(p_0,p)&=& \frac{m_D^2}{\bar{u}^2}\(1-\frac{p_0}{2p}\log\frac{p_0+p}{p_0-p}\),
\eea
which agrees with the HTL longitudinal form factor  $\Pi_L(p_0,p)$~\cite{Lebellac;1996}.
Similarly, we will calculate here the coefficients $c_0$  and $d_0$ explicitly.  
\bea
\hspace{-.5cm}c_0(p_0,p)&=&R^{\mu\nu}\Big[\Pi_{\mn}^g(P)+\Pi^{0}_{\mu\nu}(P)\Big]\nn\\
&=&  (\Pi^{g})^{\mu}_\mu(P) +(\Pi^{0})^{\mu}_\mu(P) + \frac{1}{p_\perp^2}\Big[\(p_0^2-p_\perp^2\)\Big\{\Pi_{00}^{g}(P)+\Pi_{00}^{0}(P)\Big\}\nn\\
&&+\,p^2\Big\{\Pi_{33}^{g}(P)+\Pi_{33}^{0}(P)\Big\}-2p_0p_3\Big\{\Pi_{03}^{g}(P)+\Pi_{03}^{0}(P)\Big\}\Big],
\label{c0}
\eea
and
\bea
d_0(p_0,p)&=&Q^{\mu\nu}\Big[\Pi_{\mn}^g(P)+\Pi^{0}_{\mu\nu}(P)\Big]\nn\\
&=&-\frac{p^2}{p_\perp^2}
\left[\Big\{\Pi_{33}^{g}(P)+\Pi_{33}^{0}(P)\Big\} -\frac{2p_0p_3}{p^2}\Big\{\Pi_{03}^{g}(P)+\Pi_{03}^{0}(P)\Big\}\right.\nn\\
&&\hspace{2cm}\left. +\ \frac{p_0^2p_3^2}{p^4}\Big\{\Pi_{00}^{g}(P)+\Pi_{00}^{0}(P)\Big\}\right]. \label{d0}
\eea
Now from Eq.~\eqref{Pi_mn_g}, we can write
\bea
\Pi^{g}_{00}(P)&=&\frac{N_c\,g^2T^2}{3}\(1-\frac{p_0}{2p}\log\frac{p_0+p}{p_0-p}\),\label{Pi_00_g} \\
\Pi^{g}_{03}(P)&=& \frac{N_c\,g^2T^2}{3}\frac{p_0p_3}{p^2}\(1-\frac{p_0}{2p}\log\frac{p_0+p}{p_0-p}\),\\
\Pi^{g}_{33}(P)&=&\frac{N_c\,g^2T^2}{3}\frac{3p_3^2-p^2}{p^2}\frac{p_0^2}{2p^2}\(1-\frac{p_0}{2p}\log\frac{p_0+p}{p_0-p}\) + \frac{N_c\,g^2T^2}{3}\frac{p_3^2-p^2}{2p^2}\frac{p_0}{2p}\log\frac{p_0+p}{p_0-p}. 
\eea
We note that $00$ component from the quark contribution $\Pi_{00}^{0}$ is already calculated in Eq.~(\ref{Pi00_0}) and one needs to calculate the remaining two components of $\Pi_{\mn}^0(P)$ which are as follows:
\bea
\Pi^{0}_{03}(P)&=&\sum_f\frac{ig^2}{2}\int\frac{d^4K}{(2\pi)^4}\frac{8k_0k_3}{{K^2Q^2}} =-\frac{N_fg^2T^2}{6}\int\frac{d\Omega}{4\pi}\frac{p_0\hat{k}_3}{P\cdot\hat{K}}\nn\\
&=& \frac{N_fg^2T^2}{6}\frac{p_0p_3}{p^2}\(1-\frac{p_0}{2p}\log\frac{p_0+p}{p_0-p}\), 
\label{Pi03_0}
\eea
and
\bea
\Pi^{0}_{33}(P)&=&\sum_f\frac{ig^2}{2}\int\frac{d^4K}{(2\pi)^4}\frac{8k_3^2+4K^2}{{(K^2-m_f^2)(Q^2-m_f^2)}}
=-\frac{N_fg^2T^2}{6}\int\frac{d\Omega}{4\pi}\frac{p_0\hat{k}_3^2}{P\cdot\hat{K}}\nn\\
&=&\frac{N_fg^2T^2}{6}\frac{3p_3^2-p^2}{p^2}\frac{p_0^2}{2p^2}\(1-\frac{p_0}{2p}\log\frac{p_0+p}{p_0-p}\) + \frac{N_fg^2T^2}{6}\frac{p_3^2-p^2}{2p^2}\frac{p_0}{2p}\log\frac{p_0+p}{p_0-p}.
\label{Pi33_0}
\eea
Using the results from Eqs.~(\ref{Pi00_0}), (\ref{Pi_00_g}) - (\ref{Pi33_0}), $c_0(p_0,p)$ and $d_0(p_0,p)$ become
\bea
c_0(p_0,p)=d_0(p_0,p)=\frac{m_D^2}{2p^2}\[p_0^2-\(p_0^2-p^2\)\frac{p_0}{2p}\log\frac{p_0+p}{p_0-p}\],
\eea
which agrees with the HTL transverse form factor  $\Pi_T(p_0,p)$~\cite{Lebellac;1996}.

This implies that the zero magnetic field contribution of the fourth form factor $a$ should vanish. Below we obtain the same from Eqs.~\eqref{ff_a} and \eqref{se_pi_00} as,
\bea
a_0&=&\frac{1}{2}N^{\mn}\Big[\Pi_{\mu\nu}^{g}+\Pi_{\mu\nu}^{0}\Big]\nn\\
&=&\frac{1}{2\sqrt{\bar u^2}\sqrt{\bar n^2}}\bigg[u^\mu n^\nu+n^\mu u^\nu-2
\frac{\bar u \cdot n}{\bar u^2}\bar u^\mu \bar u^\nu\bigg]\Big[\Pi_{\mu\nu}^{g}+\Pi_{\mu\nu}^{0}\Big]\nn\\
&=&\frac{1}{2\sqrt{\bar u^2}\sqrt{\bar n^2}}\bigg[-2\frac{\bar u \cdot n}{\bar u^2}\Big[\Pi_{00}^{g}+\Pi_{00}^{0}\Big]+2\Big[\Pi_{03}^{g}+\Pi_{03}^{0}\Big]\bigg]\nn\\
&=&0
\eea

\subsubsection{Computation of form factors and Debye mass of 
\texorpdfstring{$\mathcal{O}\[(eB)^2\]$}{[O(eB)2]} terms}
In this subsection, we calculate the $\mathcal{O}\[(eB)^2\]$ coefficients of $b,c,d,a$  which are denoted by $b_2,c_2,d_2,a_2$, respectively. 
The form factor $b_2$, \textit{i.e.},  $\mathcal{O}(eB)^2$ term of the coefficient $b$, has been computed in Eq.~(\ref{b2_final_app}) of appendix~\ref{b_eb_2} as 
\bea
b_2 &=&\frac{1}{\bar{u}^2}\Big[\delta\Pi_{00}^{a}(P)+2\delta\Pi_{00}^{b}(P)\Big]\nn\\
  &=& \frac{\delta m_D^2}{\bar{u}^2}+\sum_f\frac{g^2(q_fB)^2}{\bar{u}^2\pi^2}\Biggl[\left(g_k+\frac{\pi m_f-4T}{32m_f^2T}\right)(A_0-A_2)\nn\\
&&+\left(f_k+\frac{8T-\pi m_f}{128 m_f^2 T}\right)\left(\frac{5A_0}{3}-A_2\right) \Biggr] . \label{b2_final}
  \eea
 and also the Debye screening mass of $\mathcal{O}(eB)^2$  as obtained in Eq.~(\ref{debye_mass_app}) of appendix ~\ref{b_eb_2} as
 \bea
\delta m_D^2 &=&-\sum_f\frac{g^2}{3\pi^2}(q_fB)^2\Bigg[\left(\frac{\partial}{\partial (m_f^2)}\right)^2 + m_f^2\left(\frac{\partial}{\partial (m_f^2)}\right)^3 \Bigg]\nn\\
&&\times\ m_f^2
\sum\limits_{l=1}^\infty(-1)^{l+1}\left[K_2\left(\frac{m_fl}{T}\right)-K_0\left(\frac{m_fl}{T}\right)\right]\nn\\
&=&\frac{g^2}{12\pi^2T^2}\sum_f(q_fB)^2\sum\limits_{l=1}^\infty(-1)^{l+1}l^2K_0\left(\frac{m_fl}{T}\right).
\eea
We obtain $\mathcal{O}(eB)^2$ term of the coefficient $c$  in Eq.~(\ref{c2_final_app}) of appendix~\ref{c_eb_2} as 
\bea
c_2 &=& R^{\mn}(\delta\Pi_{\mn}^{a} + 2\delta\Pi_{\mn}^{b})\nn\\
&=& -\sum_f\frac{4g^2(q_fB)^2}{3\pi^2}g_k + \sum_f\frac{g^2(q_fB)^2}{2\pi^2}\left(g_k + \frac{\pi m_f - 4T}{32m_f^2T}\right) \times 
\Biggl[-\frac{7}{3} \frac{p_0^2}{p_{\perp}^2} + 
\left(2+\frac{3}{2}\frac{p_0^2}{p_{\perp}^2}\right)A_0 \nn\\
&&+\left(\frac{3}{2}+\frac{5}{2}\frac{p_0^2}{p_{\perp}^2}+\frac{3}{2}\frac{p_3^2}{p_{\perp
}^2} \right)A_2 - \frac{3p_0p_3}{p_{\perp}^2}A_1 
- \frac{5}{2}\left(1-\frac{p_3^2}{p_{\perp}^2}\right)A_4- \frac{5p_0p_3}{p_{\perp}^2}A_3  
\Biggr] . \label{c2_final}
\eea
We calculate the $\mathcal{O}(eB)^2$ term of the coefficient $d$  in appendix~\ref{d_eb_2} as
\bea
d_2 &=& Q^{\mn}(\delta\Pi_{\mn}^a + 2\delta\Pi_{\mn}^b) = F_1+F_2 , \label{final_d2}
\eea
where expressions for $F_1$ and $F_2$ can be found in Eqs.~(\ref{f1}) and (\ref{f2}), respectively.

The $\mathcal{O}(eB)^2$ term of the coefficient $a$ is calculated in appendix~\ref{a_eb_2} as
\bea
a_2 &=& N^{\mn}(\delta\Pi_{\mn}^a + 2\delta\Pi_{\mn}^b) = G_1+G_2 , \label{final_a2}
\eea
where $G_1$ and $G_2$ are given in Eqs.~(\ref{G1}) and~(\ref{G2}) respectively.

\begin{figure}[t]
{\includegraphics[scale=0.63]{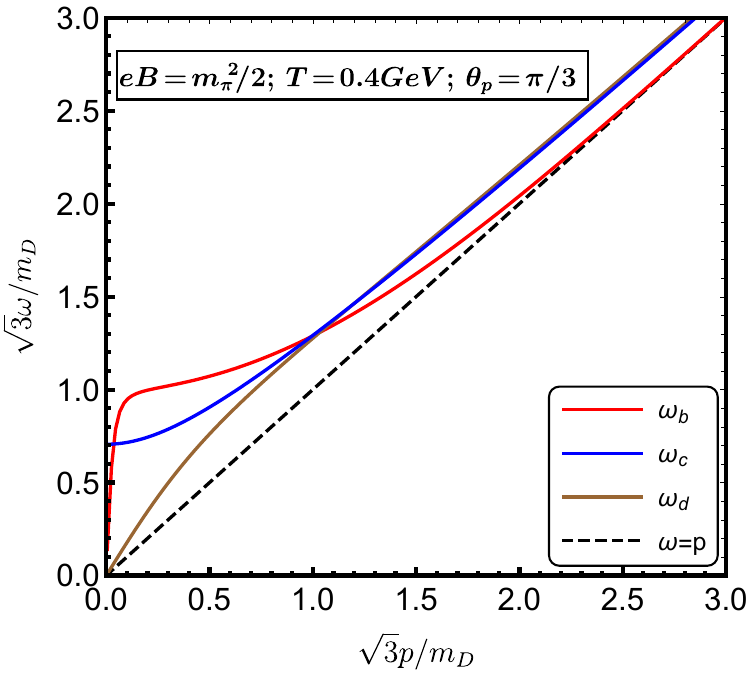}}
{\includegraphics[scale=0.63]{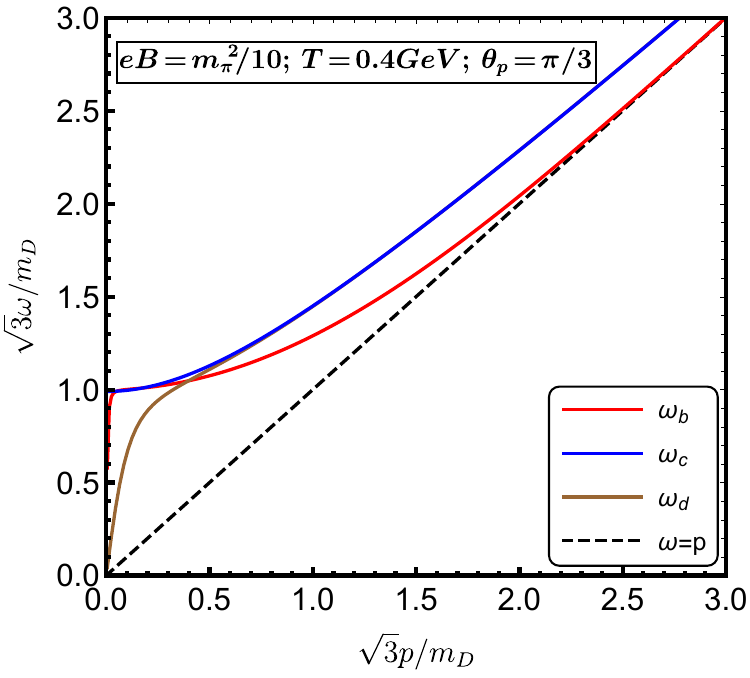}}
\hspace*{-0.5in}{\includegraphics[scale=0.7]{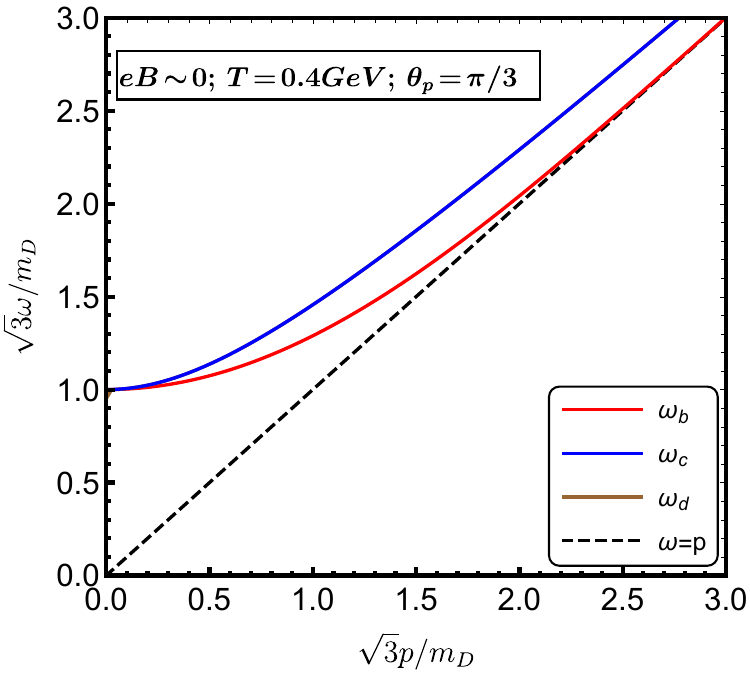}}
	\caption{Gluon dispersion curves for $\theta_p=\pi/3$ but with varying magnetic field strength 
	 $eB=m_\pi^2/2, \ \  m_\pi^2/10\ \ \rm{and} \, \  m_\pi^2/800 (\sim 0)$ for $N_f=2$.}
	\label{fig:disp_pi3}
\end{figure}

\subsubsection{Dispersion }
In weak field approximation the dispersion relation can now be written as
\begin{subequations}
\begin{align}
P^2-c=P^2-\Pi_T-c_2 & =0,\\
(P^2-b)(P^2-d)-a^2&=
(P^2-\Pi_L-b_2)(P^2-\Pi_T-d_2)-a_2^2\nn\\
&=\(P^2-\frac{b_0+b_2+d_0+d_2+\sqrt{\(b_0+b_2-d_0-d_2\)^2+4a_2^2}}{2}\)\nn\\
&\times\(P^2-\frac{b_0+b_2+d_0+d_2-\sqrt{\(b_0+b_2-d_0-d_2\)^2+4a_2^2}}{2}\)=0\label{disp_eq_abd}
\end{align}
\end{subequations}
 which give rise to $c$,\,\,$n^+$ and $n^-$ dispersive modes with energies $\omega_c$, $\omega_{n^+}$ and $\omega_{n^-}$ respectively.

In this section, we consider that the magnetic field is the smallest scale and calculate all the quantities up to $\mathcal{O}[\(eB\)^2]$. Within this approximation, Eq.~\eqref{disp_eq_abd} can be approximated as
\begin{align}
&\(P^2-b_0-b_2\)\(P^2-d_0-d_2\)=0,
\end{align}
as there is no contribution of $\mathcal O[\(eB\)^2]$ from $a_2$ and it only starts contributing $\mathcal O[\(eB\)^4]$ onwards. Thus $a_2$ can safely be neglected in the weak field 
approximation. Now one can write the dispersion relation in weak field approximation as
\bea
P^2-b &=&0,\nn\\
P^2-c&=&0,\nn\\
P^2-d &=&0,
\eea
where the respective dispersive modes are denoted by $b$-,$c$-,$d$-mode.

We note that the dispersion relations are scaled by plasma frequency of non-magnetized medium, $\omega_p=m_D/\sqrt{3}$ where $m_D^2$ is given in Eq.~\eqref{debye_mass_qed}. As seen that there are three distinct modes when a gluon propagates in hot magnetized material medium.  The magnetized plasmon mode with energy $\omega_b$ appears due to the form factor $b$  whereas two transverse modes  with energy  $\omega_c$ and $\omega_d$ are, respectively, due to the form factors $c$ and $d$. The presence of magnetic field  lifts the degeneracy of the transverse mode found only in a thermal medium.

\begin{figure}[tbh]
\subfigure{\includegraphics[scale=0.63]{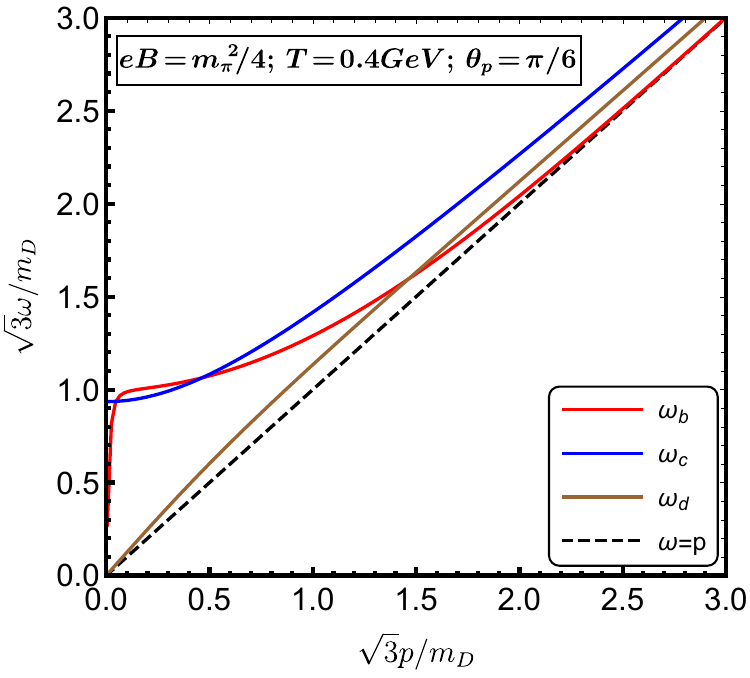}}
\subfigure{\includegraphics[scale=0.63]{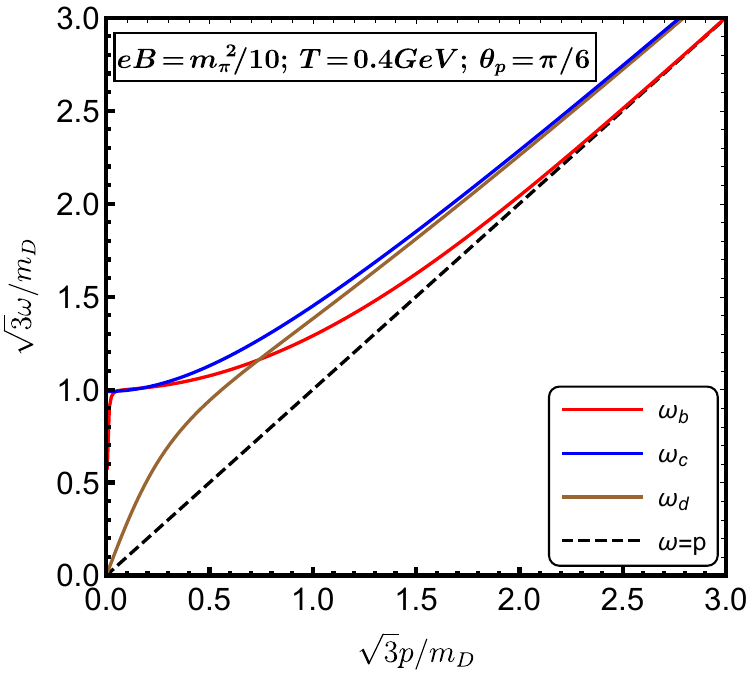}}
\hspace*{-0.5in}{\includegraphics[scale=0.7]{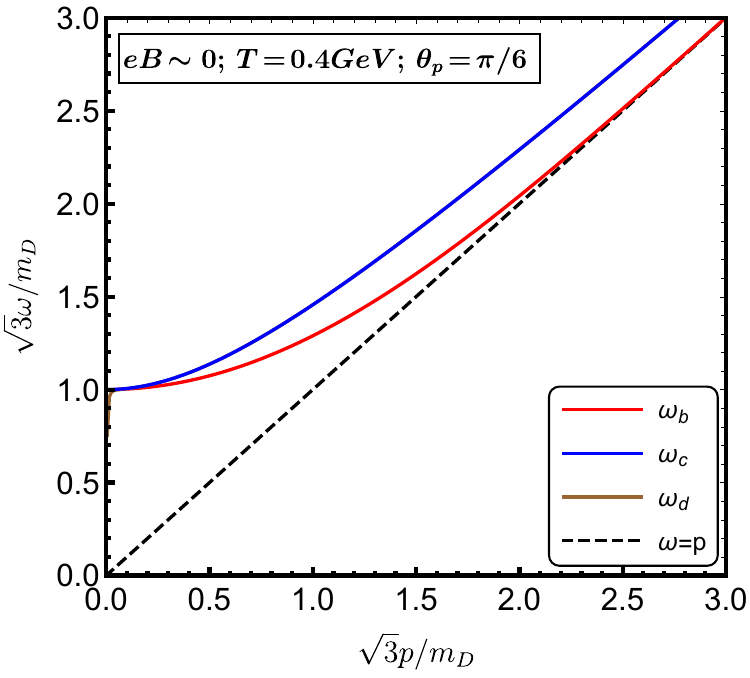}}
	\caption{Gluon dispersion curves for $\theta_p=\pi/6$ but with varying magnetic field strength 
			$eB=m_\pi^2/4, \ \  m_\pi^2/10\ \ \rm{and} \, \  m_\pi^2/800 (\sim 0)$ for $N_f=2$.}
	\label{fig:disp_pi4}
\end{figure}

Now, the dispersion curves for gluon are displayed in Fig.~\ref{fig:disp_pi3} when it propagates at an angle $\theta_p=\pi/3$ with the direction of the magnetic field.  We have chosen three different values of magnetic field $|eB|=m_\pi^2/2, \ \  m_\pi^2/10\ \ \rm{and} \, \ m_\pi^2/800 (\sim 0) $; $m_\pi$ is the pion mass.
For a given magnetic field strength,  say $|eB|=m_\pi^2/2$,  one finds two modes (viz., $b$ and $d$ mode) with vanishing plasma frequency and one mode (viz., $c$ mode)
 with finite plasma frequency. 
The zero plasma frequency for $b$ and $d$ modes could be the artefact of the weak field approximation used in the series expanded version of the Schwinger propagator, i.e. Eq.~\eqref{prop_wfa} where the propagator is expanded in a series of $eB$ by considering $eB$ as the lowest scale. This expansion constrains the arbitrariness of the value of $p$ as it is valid only when $p\gtrsim\sqrt {eB}$. Hence in the limit $p\rightarrow 0$ with finite value of $eB$ (however small), as $p$ then becomes the lowest scale and Eq.~\eqref{prop_wfa} is not valid. For $d$ mode with a very small magnetic field, the dispersion curve for $d$ at $p=0$ jumps to zero abruptly. This is because, taking $p\rightarrow 0$ limit before taking $eB \rightarrow 0$ again violates the condition $p\gtrsim\sqrt {eB}$ and leaves behind a zero frequency mode.  However, the situation is different while taking $eB\rightarrow 0$ limit first though, as in that case considering $eB=0$, one gets back
two HTL dispersive modes for gluon propagation.
In Fig.~\ref{fig:disp_pi4} we have also displayed the dispersion of gluon when it propagates at an angle $\theta_p=\pi/6$.

\section{Conclusion}
\label{conclusion}
In this article,  we have constructed the general structure of two point functions (self-energy and propagator) of a  gauge boson when it travels through a magnetized thermal medium.  The Lorentz (boost) invariance is broken due to the presence of heat bath whereas rotational invariance is broken
due to the presence of a background magnetic field. Based on gauge invariance and symmetry properties of the gauge boson self-energy, the general Lorentz structure of gauge boson
two point functions is obtained 
by using four linearly independent basis tensors.
We used the effective two point functions to study the dispersion spectra  of a gluon in hot magnetized medium.  In strong field approximation, one finds three modes which in limiting cases (propagation angle $\pi/2$) merge with the thermal modes.  On the other hand in weak field approximation one also finds three distinct modes,  {\it viz.}, one magnetized plasmon, two transverse mode.  The calculation for photon can trivially be obtained from this calculation.
We further note that the effective propagator obtained here can conveniently be used to study  various quantities in QED and QCD plasma. We, finally, note that  in a following  calculation~\cite{Bandyopadhyay:2017cle}, 
various thermodynamic quantities are computed using the general structure of the gauge boson here and fermions in Ref.~\cite{Das:2017vfh}  
of a magnetized hot  QCD plasma. 

\section{Acknowledgment}
Authors  gratefully acknowledge the valuable discussion with J. Kapusta on one important technical point in the calculation and a
 very useful discussion with Palash B Pal and Michael Strickland. The authors would also like to acknowledge the referee for numerous useful comments during the review process. BK is thankful to Arghya Mukherjee and Avik Banerjee for helpful discussions. 
 AB is supported by the National Post Doctoral Program CAPES (PNPD/CAPES), Govt. of Brazil. NH was funded by the Alexander von Humboldt Foundation, Germany as an Alexander von Humboldt  postdoctoral fellow during his stay in Germany. NH was also supported by the Department of Atomic Energy (DAE), India during his stay in India. BK and MGM were funded by the DAE, India via the project TPAES.
\appendix
\section{Notation for Frequency Sum Integral}
In imaginary time formalism  an integral over loop momentum   can be replaced  by a  frequency sum  and an integral over three momentum as
\bea
  \int \frac{d^4K}{(2\pi)^4}
   & \;\equiv\; &  \sumintb_{K}
     \;\equiv\; 
  \left(\frac{e^{\gamma_E}\Lambda^2}{4\pi}\right)^\epsilon\;
 i T\sum_{k_0=2n\pi  iT}\:\int {d^{3-2\epsilon}k \over (2 \pi)^{3-2\epsilon}}\;,\\ 
  \int \frac{d^4K}{(2\pi)^4}  &  \;\equiv\; & \sumintf_{\{K\}}
    \;\equiv\; 
  \left(\frac{e^{\gamma_E}\Lambda^2}{4\pi}\right)^\epsilon\;
  i T\sum_{k_0=(2n+1)\pi  iT}\:\int {d^{3-2\epsilon}k \over (2 \pi)^{3-2\epsilon}}\;,
\label{sumint-def}
\eea
where the loop integral is over Minkowski momentum $K$. Now, the first one is for boson whereas the second one is for fermion. 
The integral over spatial momentum, in dimensional regularization, is generalized to $d = 3-2 \epsilon$ spatial dimensions and $\Lambda$ is an arbitrary momentum scale. 
The factor $(e^{\gamma_E}/4\pi)^\epsilon$ is introduced so that, after minimal subtraction of the poles in $\epsilon$
due to ultraviolet divergences, $\Lambda$ coincides 
with the renormalization scale of the $\overline{\rm MS}$ renormalization scheme. 
\section{Calculation of \texorpdfstring{$\left (\Pi^\mu_\mu\right)^s$}{Pimumus} in strong field approximation}
\label{sfa_app}

Now combining Eqs.~(\ref{pol_vacuum} ) and (\ref{se_sfa_gen}) and then contracting with $g^{\mn}$ one can obtain 
$(\Pi_\mu^\mu)^s$  as
\bea
(\Pi_\mu^\mu)^s&=&-\sum_fe^{{-p_\perp^2}/{2|q_fB|}}~\frac{g^2|q_fB|}{2\pi}T\sum\limits_{k_0}\int \frac{dk_3}{2\pi}
\,\frac{2m_f^2}{(K_{\sp}^2-m_f^2)(Q_{\sp}^2-m_f^2)}. \label{pimm}
\eea
We note that the sum integration after Eq.(\ref{pimm}) is infrared divergent for $m_f=0$ in the limit $k_3\rightarrow0$. Below we extract the finite 
part of it using  HTL approximation and  the method used in  Ref.~\cite{Dolan:1973qd} as 
\bea
(\Pi_\mu^\mu)^s&\approx& \sum_f2m_f^2 e^{{-p_\perp^2}/{2|q_fB|}}~\frac{g^2 |q_fB|}{2\pi}\int \frac{dk_3}{2\pi}
\left[ \frac{1}{2E_{k_3}^2}\left\{\frac{n_F(E_{k_3})}{E_{k_3}} + 
 p_3 \frac{k_3}{E_{k_3}}\frac{\partial n_F(E_{k_3})}{\partial E_{k_3}}\frac{p_3 k_3/E_{k_3}}{p_0^2-p_3^2  (k_3/E_{k_3})^2}\right\}\right]\nn\\
 &\approx& \sum_f2m_f^2 e^{{-p_\perp^2}/{2|q_fB|}}~\frac{g^2 |q_fB|}{2\pi}\int \frac{dk_3}{2\pi}
 \Bigg[-\frac{\partial}{\partial (m_f^2)}\frac{n_F(E_{k_3})}{E_{k_3}}\left(1-\frac{p_0^2}{p_0^2-p_3^2}\right)+\frac{n_F(E_{k_3})}{2 E_{k_3}^3}\,\frac{p_0^2}{p_0^2-p_3^2 }\Bigg]\nn\\
 &=&\sum_f2m_f^2 e^{{-p_\perp^2}/{2 |q_fB|}}~\frac{g^2|q_fB|}{4\pi^2}\Bigg[-\frac{1}{2m_f^2}
 \left(\frac{p_3^2}{p_0^2-p_3^2}\right)\nn\\
 &&\hspace{4cm}+\left(\frac{1}{2m_f^2}-\frac{\pi}{8m_fT}+\frac{7\zeta(3)}{8\pi^2T^2}\right)\,
 \frac{p_0^2}{p_0^2-p_3^2}\Bigg] , \label{pimm_result}
 \eea 
 where we have used Eq.~\eqref{sfa_freq_sum} and the following integrals
 \bea
 \int_{-\infty}^{\infty} dk_3\,\frac{n_F(E_{k_3})}{E_{k_3}}&=&-\log{\frac{m_f}{\pi T}}-\gamma_E,\nn\\
 -\frac{\partial}{\partial (m_f^2)}\int_{-\infty}^{\infty} dk_3\,\frac{n_F(E_{k_3})}{E_{k_3}}&=&\frac{1}{2m_f^2} , \nn\\
  \int_{-\infty}^{\infty} dk_3\,\frac{n_F(E_{k_3})}{2E_{k_3}^3}&=&\frac{1}{2m_f^2}-\frac{\pi}{8m_fT}+\frac{7\zeta(3)}{8\pi^2T^2} .
  \eea
  Now, using Eq.~(\ref{d_m_sf}) we have
  \bea
   d&=&\sum_f2m_f^2~e^{{-p_\perp^2}/{2 |q_fB|}}~\frac{g^2  |q_fB|}{4\pi^2}\Bigg[-\frac{1}{2m_f^2}
 \left(\frac{p_3^2}{p_0^2-p_3^2}\right)\nn\\
 &&\!\!+\left(\frac{1}{2m_f^2}-\frac{\pi}{8m_fT}+\frac{7\zeta(3)}{8\pi^2T^2}\right)\,
 \frac{p_0^2}{p_0^2-p_3^2}\Bigg]+\sum_f e^{{-p_\perp^2}/{2  |q_fB|}}\!\left(\frac{\delta m_{D,f}}{\bar u}\right)^2\!
 \frac{p_3^2}{p_0^2-p_3^2} . \label{d_sf}
 \eea  
\section{Simplification using HTL  approximation}
\label{htl}
Now, based on HTL approximation we simplify the  terms in Eq.~(\ref{b1_wfa})  as 
\begin{subequations}
\begin{align}
&\hspace{-1.5cm}\frac{\partial}{\partial\!\(m_f^2\)}\frac{1}{(K^2-m_f^2)^2(Q^2-m_f^2)}\nn\\&=\frac{2}{(K^2-m_f^2)^3(Q^2-m_f^2)}+\frac{1}{(K^2-m_f^2)^2(Q^2-m_f^2)^2}\nn\\
&\simeq \frac{3}{(K^2-m_f^2)^3(Q^2-m_f^2)},\label{approx1} \\
&\hspace{-1.5cm}\frac{\partial}{\partial\!\(m_f^2\)}\frac{1}{(K^2-m_f^2)^3(Q^2-m_f^2)}\nn\\
&=\frac{3}{(K^2-m_f^2)^4(Q^2-m_f^2)}+\frac{1}{(K^2-m_f^2)^3(Q^2-m_f^2)^2}\nn\\
&\simeq \frac{4}{(K^2-m_f^2)^4(Q^2-m_f^2)},\label{approx2} \\
&\hspace{-1.5cm}\frac{\partial}{\partial\!\(m_f^2\)}\frac{1}{(K^2-m_f^2)^n(Q^2-m_f^2)}\nn\\&\approx \frac{\partial}{\partial\!\(k^2\)}\frac{1}{(K^2-m_f^2)^n(Q^2-m_f^2)} = \frac{1}{2k}\frac{\partial}{\partial k}\frac{1}{(K^2-m_f^2)^n(Q^2-m_f^2)},\label{approx3} \\
&\hspace{-1.5cm}\int\frac{d^4K}{(2\pi)^4}\frac{1}{(K^2-m_f^2)^2(Q^2-m_f^2)}\nn\\
&= -2 \int\frac{d^4K}{(2\pi)^4}\frac{k^2}{(K^2-m_f^2)^3(Q^2-m_f^2)}\nn\\
&= \frac{16}{5}\int\frac{d^4K}{(2\pi)^4}\frac{k^4}{(K^2-m_f^2)^4(Q^2-m_f^2)}, \label{approx4} \\
&\hspace{-1.5cm}\int\frac{d^4K}{(2\pi)^4}\frac{1}{(K^2-m_f^2)^3(Q^2-m_f^2)}\nn\\
&= -\frac{8}{3} \int\frac{d^4K}{(2\pi)^4}\frac{k^2}{(K^2-m_f^2)^4(Q^2-m_f^2)}.\label{approx5}
\end{align}
\end{subequations}
\section{Frequency sum}
\label{freq_sum}
\bea
&&\hspace{-1cm}i\int\frac{d^4K}{(2\pi)^4}\frac{1}{(K^2-m_f^2)(Q^2-m_f^2)} \nn\\
&=& \int\frac{k^2\,dk}{2\pi^2}\int\frac{d\Omega}{4\pi}\,\frac{1}{2E_k^2}\left[\frac{n_F(E_k)}{E_k} + 
\frac{\partial n_F(E_k)}{\partial k}\frac{\bm{p\cdot\hat k}}{p_0-\bm{p\cdot\hat k}\,k/E_k}\right]\nn\\
&=& \int\frac{k^2\,dk}{2\pi^2}\int\frac{d\Omega}{4\pi}\,\frac{1}{2E_k^2}\left[\frac{n_F(E_k)}{E_k} + 
\frac{\partial n_F(E_k)}{\partial E_k}\frac{\bm{p\cdot\hat k}}{p_0 E_k/k-\bm{p\cdot\hat k}}\right]\nn\\
&\approx& \int\frac{k^2\,dk}{2\pi^2}\int\frac{d\Omega}{4\pi}\frac{1}{2E_k^2}\left[\frac{n_F(E_k)}{E_k} + 
\frac{\partial n_F(E_k)}{\partial E_k}\,\frac{\bm{p\cdot\hat k}}{p_0-\bm{p\cdot\hat k}}\right]\nn\\
&=& \int\frac{k^2\,dk}{2\pi^2}\int\frac{d\Omega}{4\pi}\left[\frac{1}{2E_k^2}\frac{n_F(E_k)}{E_k} - 
\frac{1}{2E_k^2} \frac{\partial n_F(E_k)}{\partial E_k} + \frac{1}{2E_k^2}\frac{\partial n_F(E_k)}{\partial E_k} \frac{p_0}{p_0 - \bm{p\cdot\hat k}}\right]\nn\\
&=& \int\frac{k^2\,dk}{2\pi^2}\int\frac{d\Omega}{4\pi}\,\left[-\frac{\partial}{\partial\!\(E_k^2\)}\frac{n_F(E_k)}{E_k} + \frac{1}{E_k}\frac{\partial n_F(E_k)}{\partial\!\(E_k^2\)}\frac{p_0}{P\cdot \hat K}\right]\nn\\
&=& \int\frac{k^2\,dk}{2\pi^2}\int\frac{d\Omega}{4\pi}\left[-\frac{\partial}{\partial\!\(m_f^2\)}\frac{n_F(E_k)}{E_k} + \frac{1}{E_k}\frac{\partial n_F(E_k)}{\partial\!\(m_f^2\)}\frac{p_0}{P\cdot \hat K}\right]\nn\\
&=& \int\frac{k^2\,dk}{2\pi^2}\int\frac{d\Omega}{4\pi}\left[-\frac{\partial}{\partial\!\(m_f^2\)}\frac{n_F(E_k)}{E_k} + \frac{\partial}{\partial\!\(m_f^2\)}\frac{n_F(E_k)}{E_k}\,\frac{p_0}{P\cdot \hat K} + \frac{n_F(E_k)}{2 E_k^3}\frac{p_0}{P\cdot \hat K}\right]\nn\\
&=& -\frac{\partial}{\partial\!\(m_f^2\)}\!\int\!\frac{k^2\,dk}{2\pi^2}\frac{n_F(E_k)}{E_k}\!\int\!\frac{d\Omega}{4\pi}\left[1-\frac{p_0}{P\cdot \hat K}\right] + \!\int\!\frac{k^2dk}{2\pi^2}\!\int\!\frac{d\Omega}{4\pi}\,\frac{n_F(E_k)}{2 E_k^3}\frac{p_0}{P\cdot \hat K}
\eea
Let us take $m_f=yT$ and $k=xT$.
\be
\frac{\partial}{\partial\!\(m_f^2\)}\int k^2\,dk\,\frac{n_F\!\(\sqrt{k^2+m_f^2}\)}{\sqrt{k^2+m_f^2}}=\frac{\partial}{\partial\!\(y^2\)}\int x^2\,dx\,\frac{n_F\!\(\sqrt{x^2+y^2}\)}{\sqrt{x^2+y^2}}
\ee
The integrals can be represented by the well-known functions as,
\be
f_{n+1}(y)=\frac{1}{\Gamma(n+1)}\int_0^\infty\frac{dx\,x^n}{\sqrt{x^2+y^2}}n_F\(\sqrt{x^2+y^2}\)
\ee
which satisfy the following recursion relation,
\be
\frac{\partial f_{n+1}}{\partial y^2} = -\frac{f_{n-1}}{2n}
\ee
In the regime of HTL perturbation theory and weak magnetic field, one can use high temperature expansion for $f_1$ as,
\be
f_1 = -\frac{1}{2}\ln{(y/\pi)}-\frac{1}{2}\gamma_E
\ee
\be
\mbox{So, }\frac{\partial}{\partial\!\(m_f^2\)}\int\frac{k^2\,dk}{2\pi^2}\frac{n_F(E_k)}{E_k} = \frac{1}{8\pi^2}\left[\ln{\frac{m_f}{\pi T}}+\gamma_E\right]
\ee
 \be
  \int\limits_0^\infty\frac{dk\,k^2}{2\pi^2}\frac{n_F(E_k)}{2E_k^3}\!\int\!\frac{d\Omega}{4\pi}\frac{p_0}{P\cdot \hat K} = -\frac{1}{8\pi^2}\left[1+\gamma_E-\frac{m_f\pi}{4T}+\ln{\frac{m_f}{\pi T}}\right]\int\frac{d\Omega}{4\pi}\frac{p_0}{P\cdot \hat K},
 \ee
 \bea
 &&\hspace{-1cm}\int\frac{d^4K}{(2\pi)^4}\,\frac{1}{(K^2-m_f^2)(Q^2-m_f^2)} \\
 &=& -\frac{1}{8\pi^2}\left[\ln{\frac{m_f}{\pi T}}+\gamma_E\right]\int\frac{d\Omega}{4\pi}\left[1-\frac{p_0}{P\cdot \hat K}\right]\nn\\
 &&-\frac{1}{8\pi^2}\left[1+\gamma_E-\frac{m_f\pi}{4T}+\ln{\frac{m_f}{\pi T}}\right]\int\frac{d\Omega}{4\pi}\frac{p_0}{P\cdot \hat K}\nn\\
 &=& -\frac{1}{8\pi^2}\left[\ln{}\frac{m_f}{\pi T}+\gamma_E\right] + \frac{1}{8\pi^2}\left[\frac{m\pi}{4 T} - 1\right]\int\frac{d\Omega}{4\pi}\frac{p_0}{P\cdot \hat K}\nn\\
 &=& \frac{1}{8\pi^2}\left[-\ln{\frac{m_f}{\pi T}}-\gamma_E+\left(\frac{m_f \pi}{4 T}-1\right)\int\frac{d\Omega}{4\pi}\frac{p_0}{P\cdot \hat K}\right].
 \eea
\bea
i\int \frac{d^4K}{(2\pi)^4}\frac{k_0 k c}{(K^2-m_f^2)(Q^2-m_f^2)}&=& -\int \frac{k^2 dk}{2\pi^2}k c \frac{\partial n_F(E_k)}{\partial(m_f^2)}\int \frac{d\Omega}{4\pi}\bigg(1-\frac{p_0}{P \cdot \hat K}\bigg),\\
i\int\frac{d^2K_{\sp}}{(2\pi)^2}\frac{1}{(K_{\sp}^2-m_f^2)(Q_{\sp}^2-m_f^2)}&=&  \int_{-\infty}^{\infty} \frac{dk_3}{2\pi}
 \Bigg[\frac{\partial}{\partial (m_f^2)}\frac{n_F(E_{k_3})}{E_{k_3}}\frac{p_3^2}{p_0^2-p_3^2}+\frac{n_F(E_{k_3})}{2 E_{k_3}^3}\,\frac{p_0^2}{p_0^2-p_3^2 }\Bigg], \label{sfa_freq_sum}\\
i\int\frac{d^2K_{\sp}}{(2\pi)^2}\frac{k_0 k_3}{(K_{\sp}^2-m_f^2)(Q_{\sp}^2-m_f^2)}&=&\int_{-\infty}^{\infty}\frac{dk_3}{2\pi} \frac{p_0 p_3 k_3^2}{2E_{k_3}^2} \frac{\partial n_F(E_{k_3})}{\partial E_{k_3}} \,\frac{1}{p_0^2-p_3^2k_3^2/E_{k_3}^2}.
\eea
\section{Calculation of the form factors in weak field approximation}
\subsection{Calculation of the form factor \texorpdfstring{$b_2$}{b2} }
\label{b_eb_2}
\bea
b_2 &=&\frac{1}{\bar{u}^2}\Big[\delta\Pi_{00}^{a}(P)+2\delta\Pi_{00}^{b}(P)\Big]\nn\\
&=& \sum_f\frac{2i 
g^2(q_fB)^2}{\bar{u}^2}\int\frac{d^4K}{(2\pi)^4}\Biggl[\frac{K^2+(1+c^2)k^2+m_f^2}{(K^2-m_f^2)^2(Q^2-m_f^2)^2}+\frac{8(
K^2+k^2)}{(K^2-m_f^2)^3(Q^2-m_f^2)}\nn\\
&&-\frac{8(K^2+k^2)(K^2+(1-c^2)k^2-m_f^2)}{(K^2-m_f^2)^4(Q^2-m_f^2)}\Biggr],
\label{b1_wfa}
\eea
where we write $k_3$ as $ck$ with $c=\cos\theta$.
Using   Eqs.~(\ref{approx1}), (\ref{approx2}), (\ref{approx3}), and (\ref{approx4}) 
 obtained in appendix~\ref{htl} within HTL approximation,  Eq.~(\ref{b1_wfa}) becomes
\bea
b_2 &=&\sum_f\frac{2ig^2(q_fB)^2}{\bar{u}^2}\Biggl[\left(\frac{\partial}{\partial\!\( 
m_f^2\)}+\frac{m_f^2}{2}\frac{\partial^2}{\partial\!\(m_f^2\)^2}\right)
\int\frac{d^4K}{(2\pi)^4}\frac{(1-c^2)}{(K^2-m_f^2)(Q^2-m_f^2)}\nn\\
&&+\left(\frac{m_f^2}{3}\frac{\partial^2}{\partial\!\(m_f^2\)^2}\right)
\int\frac{d^4K}{(2\pi)^4}\frac{1}{(K^2-m_f^2)(Q^2-m_f^2)}\Biggr]\nn\\
&=&\frac{4i(e^2B)^2}{\bar{u}^2}\Biggl[\left(\frac{\partial}{\partial\!\( 
m_f^2\)}+\frac{5m_f^2}{6}\frac{\partial^2}{\partial\!\(m_f^2\)^2}\right)
\int\frac{d^4K}{(2\pi)^4}\frac{1}{(K^2-m_f^2)(Q^2-m_f^2)}\nn\\
&&-\left(\frac{\partial}{\partial 
m_f^2}+\frac{m_f^2}{2}\frac{\partial^2}{\partial\!\(m_f^2\)^2}\right)
\int\frac{d^4K}{(2\pi)^4}\frac{c^2}{(K^2-m_f^2)(Q^2-m_f^2)}\Biggr].
\eea
After performing the frequency sum as given in appendix~\ref{freq_sum}, we obtain
\bea
b_2 &=&\sum_f\frac{g^2(q_fB)^2}{\bar{u}^2\pi^2}\Biggl[\left(\frac{\partial^2}{\partial\!\(
m_f^2\)^2}+\frac{5m_f^2}{6}\frac{\partial^3}{\partial\!\(m_f^2\)^3}\right)
\int k^2 dk\frac{n_F(E_k)}{E_k}\int\frac{d\Omega}{4\pi}\left[\frac{p_0}{P\cdot \hat{K}}-1 \right]\nn\\
&&+\left(\frac{\partial}{\partial\!\(
m_f^2\)}+\frac{5m_f^2}{6}\frac{\partial^2}{\partial\!\(m_f^2\)^2}\right)
\int k^2 dk\frac{n_F(E_k)}{2E_k^3}\int\frac{d\Omega}{4\pi}\frac{p_0}{P\cdot \hat{K}}\nn\\
&& -\left(\frac{\partial^2}{\partial\!\(
m_f^2\)^2}+\frac{m_f^2}{2}\frac{\partial^3}{\partial\!\(m_f^2\)^3}\right)
\int k^2 dk\frac{n_F(E_k)}{E_k}\int\frac{d\Omega}{4\pi} \, c^2\left[\frac{p_0}{P\cdot \hat{K}}-1 \right]\nn\\
&& - \left(\frac{\partial}{\partial\!\(
m_f^2\)}+\frac{m_f^2}{2}\frac{\partial^2}{\partial\!\(m_f^2\)^2}\right)
\int k^2 dk\frac{n_F(E_k)}{2E_k^3}\int\frac{d\Omega}{4\pi}\, c^2\frac{p_0}{P\cdot \hat{K}}\Biggr]\nn\\
&=& \frac{3}{2} \frac{\delta m_D^2}{\bar{u}^2} \int\frac{d\Omega}{4\pi} (1-c^2)\left[1-\frac{p_0}{P\cdot \hat{K}}\right]\nn\\
&+&\sum_f\frac{g^2(q_fB)^2}{\bar{u}^2\pi^2}\Biggl\{\left(\frac{\partial}{\partial\!\(
m_f^2\)}+m_f^2\frac{\partial^2}{\partial\!\(m_f^2\)^2}\right)
\int k^2 dk\frac{n_F(E_k)}{2E_k^3}\int\frac{d\Omega}{4\pi} (1-c^2) \frac{p_0}{P\cdot \hat{K}}\nn\\
&+&  \frac{m_f^2}{2}\frac{\partial^3}{\partial\!\(m_f^2\)^3}
\int k^2 dk\frac{n_F(E_k)}{E_k}\int\frac{d\Omega}{4\pi} \left(\frac{1}{3}-c^2\right)\left[1-\frac{p_0}{P\cdot 
\hat{K}}\right]\nn\\
&-&  \frac{m_f^2}{2}\frac{\partial^2}{\partial\!\(m_f^2\)^2}
\int k^2 dk\frac{n_F(E_k)}{2E_k^3} \int\frac{d\Omega}{4\pi} \left(\frac{1}{3}-c^2\right)\frac{p_0}{P\cdot \hat{K}}\Biggr\},
\label{b2_before_kint}
\eea
where in the second line we have rearranged the terms after using the  expression of $\delta m_D^2$ a obtained following Eq.~(\ref{debye_mass}) as
\bea
\delta m_D^2 &=& \bar{u}^2 b_2 \vert_{p_0=0,p \rightarrow 0} = \Big[\delta\Pi_{00}^{a}(P)+2\delta\Pi_{00}^{b}(P)\Big]_{p_0=0,p \rightarrow 0} \nn\\
&=& -\sum_f\frac{g^2(q_fB)^2}{\pi^2}\left[\frac{2}{3}\frac{\partial^2}{\partial\!\(m_f^2\)^2} + 
\frac{2}{3}m_f^2\frac{\partial^3}{\partial\!\(m_f^2\)^3} \right]\int k^2 dk\frac{n_F(E_k)}{E_k}\nn\\
&=& -\sum_f\frac{2g^2}{3\pi^2}(q_fB)^2\left[\frac{\partial^2}{\partial\!\(m_f^2\)^2} + m_f^2\frac{\partial^3}{\partial\!
\(m_f^2\)^3} 
\right]\int k^2 dk\frac{n_F(E_k)}{E_k}.
\label{deltamD_before_kint}
\eea
There are two types of integrations that appear in Eqs.~(\ref{b2_before_kint}) and~(\ref{deltamD_before_kint}), namely, 
 \begin{subequations}
 	\begin{align}
 	I_1=\int k^2 dk\frac{n_F(E_k)}{E_k},\label{I1}\\
 	I_2=\int k^2 dk\frac{n_F(E_k)}{E_k^3}.\label{I2}
 \end{align}
\end{subequations}
Eq.~(\ref{I1}) can be evaluated in terms of Bessel function as done in Ref.~\cite{Alexandre:2000jc} and can be obtained as
\bea
I_1&=&\sum_{l=1}^{\infty}(-1)^{l+1}\int\limits_0^\infty\frac{k^2dk}{\sqrt{k^2+m_f^2}}~~e^{-\frac{\left(\sqrt{k^2+m_f^2}\right) l}{T}}\nn\\
&=&\sum_{l=1}^{\infty}(-1)^{l+1}\frac{m_f^2}{2}\left[K_2\left(\frac{m_fl}{T}\right)-K_0\left(\frac{m_fl}{T}\right)\right].
\label{I1_final}
\eea
The second integral in Eq.~(\ref{I2}) can be evaluated using the procedure described in Ref.~\cite{Dolan:1973qd} and can be obtained at small quark mass as
\bea
I_2=-\frac{1}{2}\Big[1+\gamma_E -\frac{\pi m_f}{4T}+\log\frac{ m_f}{\pi T}\Big].
\label{I2_final}
\eea 
Now, using the Eq.~(\ref{I1_final}), Eq.~(\ref{deltamD_before_kint}) can be written as
\bea
\delta m_D^2 &=&-\sum_f\frac{g^2}{3\pi^2}(q_fB)^2\Bigg[\frac{\partial^2}{\partial\!\(m_f^2\)^2} + m_f^2\frac{\partial^3}{\partial\!\(m_f^2\)^3} \Bigg]\nn\\
&&\times\ m_f^2
\sum\limits_{l=1}^\infty(-1)^{l+1}\left[K_2\left(\frac{m_fl}{T}\right)-K_0\left(\frac{m_fl}{T}\right)\right]\nn\\
&=&\sum_f\frac{g^2}{12\pi^2 T^2}(q_fB)^2\sum\limits_{l=1}^\infty(-1)^{l+1}l^2K_0\left(\frac{m_fl}{T}\right), \label{debye_mass_app}
\eea
which agrees with Ref.~\cite{Alexandre:2000jc}.

Now, we can calculate all the $k$-integrations that appear in Eq.~(\ref{b2_before_kint}) using the Eqs.~(\ref{I1_final}) and~(\ref{I2_final}) as
\bea
\left(\frac{\partial}{\partial
	\!\(m_f^2\)}+m_f^2\frac{\partial^2}{\partial\!\(m_f^2\)^2}\right)
\int k^2 dk\frac{n_F(E_k)}{2E_k^3} &=& \frac{\pi}{64 T m_f},\nn\\
\frac{m_f^2}{2}\frac{\partial^2}{\partial\!\(m_f^2\)^2}
\int k^2 dk\frac{n_F(E_k)}{2E_k^3} &=& \frac{8T-\pi m_f}{128 T m_f^2},\nn\\
\frac{m_f^2}{2}\frac{\partial^3}{\partial\!\(m_f^2\)^3}
\int k^2 dk\frac{n_F(E_k)}{E_k} = f_k &=&- \sum\limits_{l=1}^{\infty}(-1)^{l+1} \frac{l^2}{16T^2} K_2 \left(\frac{m_f 
	l}{T}\right), \nn\\
\frac{\partial^2}{\partial\!\(m_f^2\)^2}\int k^2\,dk\frac{n_F}{E_k} =g_k&=& \sum\limits_{l=1}^{\infty}(-1)^{l+1} \frac{l}{4m_fT} K_1 \left(\frac{m_f 
	l}{T}\right). \label{gk}
\eea

Next, we have to evaluate all the angular integrals of Eq.~(\ref{b2_before_kint}). The results are given below,
\bea
\int\frac{d\Omega}{4\pi} (1-c^2)\left[1-\frac{p_0}{P\cdot \hat{K}}\right] &=& \frac{2}{3} - A_0 + A_2 ,\nn\\
\int\frac{d\Omega}{4\pi} (1-c^2) \frac{p_0}{P\cdot \hat{K}} &=& A_0 - A_2 ,\nn\\
\int\frac{d\Omega}{4\pi} \left(\frac{1}{3}-c^2\right)\left[1-\frac{p_0}{P\cdot 
\hat{K}}\right] &=& -\frac{A_0}{3} + A_2, \nn\\
\int\frac{d\Omega}{4\pi} \left(\frac{1}{3}-c^2\right)\frac{p_0}{P\cdot 
\hat{K}} &=& \frac{A_0}{3} - A_2,
\eea
where $A_n$ is defined as
\bea
A_n &=& \int\frac{d\Omega}{4\pi} \frac{p_0c^n}{P\cdot \hat{K}}. \label{an}
\eea
$A_0$ and $A_2$ can now be evaluated as
\bea
A_0 &=& \int\frac{d\Omega}{4\pi} \frac{p_0}{P\cdot \hat{K}} = \frac{p_0}{2p} \log\left(\frac{p_0+p}{p_0-p}\right),\nn\\
A_2 &=& \int\frac{d\Omega}{4\pi} \frac{c^2p_0}{P\cdot \hat{K}}\nn\\
&=& \frac{p_0^2}{2p^2}\(1-\frac{3p_3^2}{p^2}\)\(1-\frac{p_0}{2p}\log\frac{p_0+p}{p_0-p}\) + \frac{1}{2}\(1-\frac{p_3^2}{p^2}\)\frac{p_0}{2p}\log\frac{p_0+p}{p_0-p}.\,
\eea
Incorporating all these we finally obtain
\bea
b_2 &=& \frac{\delta m_D^2}{\bar{u}^2}+\sum_f\frac{g^2(q_fB)^2}{\bar{u}^2\pi^2}\Biggl[\left(g_k+\frac{\pi m_f-4T}{32m_f^2T}\right)(A_0-A_2)\nn\\
&&+\left(f_k+\frac{8T-\pi m_f}{128 m_f^2 T}\right)\left(\frac{5A_0}{3}-A_2\right) \Biggr] . \label{b2_final_app}
\eea
\subsection{Calculation of the form factor \texorpdfstring{$c_2$}{c2} }
\label{c_eb_2}
In this appendix we calculate the $\mathcal{O}(eB)^2$ term of the coefficient $c$ as
\bea
c_2 &=& R^{\mn}(\delta\Pi_{\mn}^{a} + 2\delta\Pi_{\mn}^{b})\nn\\
&=& \sum_f\frac{ig^2(q_fB)^2}{2}\,\int \frac{d^4K}{(2\pi)^4}\,\Bigg[\frac{4k_0^2-4k_3^2-4m_f^2}{(K^2-m_f^2)^2(Q^2-m_f^2)^2} + \frac{4(4k_3^2-4k_0^2+4m_f^2)}{(K^2-m_f^2)^3(Q^2-m_f^2)}\nn\\
&&-\frac{4(k_0^2-k_3^2-m_f^2)(8k_\perp^2-4K^2+4m_f^2+8(\bm{k}\cdot \bm{p})_\perp^2/p_\perp^2}{(K^2-m_f^2)^4(Q^2-m_f^2)}\Bigg]\nn\\
&=& \sum_f2ig^2(q_fB)^2\,\int \frac{d^4K}{(2\pi)^4}\,\Bigg[\frac{1}{(K^2-m_f^2)^2(Q^2-m_f^2)} - \frac{k^2(1-\cos^2{\theta})\cos^2{\phi}}{(K^2-m_f^2)^3(Q^2-m_f^2)} \nn\\
&&-\frac{7k^2(1-\cos^2{\theta})(1+\cos^2{\phi})}{(K^2-m_f^2)^3(Q^2-m_f^2)} - \frac{8k^4\sin^4{\theta}(1+\cos^2{\phi})}{(K^2-m_f^2)^4(Q^2-m_f^2)}\Bigg].
\label{c2_ini}
\eea
Now, applying HTL approximations, Eq.~(\ref{c2_ini}) can be simplified as
\bea
c_2 &=& \sum_f 2ig^2(q_fB)^2\,\int \frac{d^4K}{(2\pi)^4}\,\Bigg[\frac{1}{2} + \frac{1}{4}(1-\cos^2{\theta})\cos^2{\phi} + \frac{7}{4}\sin^2{\theta}(1+\cos^2{\phi})\nn\\
&&\hspace{3cm} -\frac{5}{4}\sin^4{\theta}(1+\cos^2{\phi})\Bigg]\frac{\partial}{\partial (m_f^2)}\,\frac{1}{(K^2-m_f^2)(Q^2-m_f^2)}\nn\\
&=&\sum_f 2ig^2(q_fB)^2\,\int \frac{d^4K}{(2\pi)^4}\,\Bigg[\frac{1}{2} + 2\sin^2{\theta}\cos^2{\phi} + \frac{7}{4}\sin^2{\theta}\nn\\
&&\hspace{3cm} -\frac{5}{4}\sin^4{\theta}(1+\cos^2{\phi})\Bigg]\frac{\partial}{\partial (m_f^2)}\,\frac{1}{(K^2-m_f^2)(Q^2-m_f^2)}\nn\\
 &=& -\sum_f\frac{4g^2(q_fB)^2}{3\pi^2}g_k + \frac{g^2(q_fB)^2}{2\pi^2}\left(g_k + \frac{\pi m_f - 4T}{32m_f^2T}\right) \times 
\Biggl[-\frac{7}{3} \frac{p_0^2}{p_{\perp}^2} + 
\left(2+\frac{3}{2}\frac{p_0^2}{p_{\perp}^2}\right)A_0 \nn\\
&&+\left(\frac{3}{2}+\frac{5}{2}\frac{p_0^2}{p_{\perp}^2}+\frac{3}{2}\frac{p_3^2}{p_{\perp
} ^2 } \right)A_2 - \frac{3p_0p_3}{p_{\perp}^2}A_1 
- \frac{5}{2}\left(1-\frac{p_3^2}{p_{\perp}^2}\right)A_4- \frac{5p_0p_3}{p_{\perp}^2}A_3  
\Biggr] . \label{c2_final_app}
\eea
where $g_k$ is given in Eq.~(\ref{gk}) and  $A_1$, $A_3$  and $A_4$  are obtained using Eq.~(\ref{an}) as
\bea
A_1 &=& \int\frac{d\Omega}{4\pi} \frac{cp_0}{P\cdot \hat{K}} =
-\frac{p_0p_3}{p^2}\left[1-\frac{p_0}{2p}
\log\left(\frac{p_0+p}{p_0-p}\right)\right],\nn\\
A_3 &=& \int\frac{d\Omega}{4\pi} \frac{c^3p_0}{P\cdot \hat{K}} =
\frac{p_0}{2p}\frac{p_3}{p}\(1-\frac{5}{3}\frac{p_3^2}{p^2}\)\nn\\
&&\hspace{2cm}-\frac{3}{2}\frac{p_0}{p}\frac{p_3}{p}\(1-\frac{p_0^2}{p^2}-\frac{p_3^2}{p^2}+\frac{5}{3}\frac{p_0^2}{p^2}\frac{p_3^2}{p^2}\)\(1-\frac{p_0}{2p}\log\frac{p_0+p}{p_0-p}\),\nn\\
A_4 &=&\! \int\frac{d\Omega}{4\pi} \frac{c^4p_0}{P\cdot\hat{K}} =
\frac{3}{8}\left(1-\frac{p_3^2}{p^2}\right)^2-\frac{p_0^2}{8p^2}\left(1-\frac{5p_3^2}{p^2}\right)^2+\frac{5}{3}\frac{p_0^2}{p^2}\frac{p_3^4}{p^4}\nn\\
&-&\frac{3}{8}\left\{\!\left(1-\frac{p_0^2}{p^2}\right)^2
-\frac{2p_3^2}{p^2} \left(1-\frac{3p_0^2}{p^2}\right)^2+\frac{p_3^4}{p^4} \left(1-\frac{5p_0^2}{p^2}\right)^2+\frac{8p_0^4}{p^4}\frac{p_3^2}{p^2} \left(1-\frac{5p_3^2}{3p^2}\right)\!\right\}\! \nn\\
&\times&\left(1-\frac{p_0}{2 p} \log\frac{p_0+p}{p_0-p}\right).
\eea
\subsection{Calculation of the form factor \texorpdfstring{$d_2$}{d2}}

\label{d_eb_2}
In this appendix we compute the form factor  $d_2$ as 
\bea
d_2 &=& Q^{\mn}(\delta\Pi_{\mn}^a + 2\delta\Pi_{\mn}^b)\nn\\
&=& -\sum_f\frac{2ig(q_fB)^2p^2}{p_{\perp}^2}\int\frac{d^4K}{(2\pi)^4}\Bigg[\frac{k_0^2+k_3^2-m_f^2-\frac{4p_0p_3}{p^2}k_0k_3+\frac{p_0^2p_3^2}{p^4}(k_0^2+k_3^2+m_f^2)}{(K^2-m_f^2)^2(Q^2-m_f^2)^2}\nn\\
&+&4\Bigg(\frac{k_3^2+k_0^2-m_f^2-\frac{4p^0p^3}{p^2}k^0k^3+\frac{p_0^2p_3^2}{p^4}(k_0^2+k_3^2+m_f^2)}{(K^2-m_f^2)^3(Q^2-m_f^2)}\nn\\
&-&\frac{(k_0^2-k_3^2-m_f^2)(2k_3^2+K^2-m_f^2-\frac{4p^0p^3}{p^2}k^0k^3+\frac{p_0^2p_3^2}{p^4}(2k_0^2-K^2+m_f^2))}{(K^2-m_f^2)^4(Q^2-m_f^2)}\Bigg)\Bigg]\nn\\
&=& \sum_f\frac{2g^2(q_fB)^2p^2}{p_{\perp}^2}\!\int\! \frac{k^2dk}{2\pi^2}\Bigg[\left\{\frac{1}{4}-\(\frac{3}{2}+\frac{p_0^2p_3^2}{p^4}\)c^2+\frac{5}{4}c^4\right\}
\frac{\partial}{\partial\!\(m_f^2\)}+m_f^2\frac{p_0^2p_3^2}{2p^4}(5-c^2)\frac{\partial^2}{\partial\!\(m_f^2\)^2}\Bigg]\nn\\
&\times&\left\{\frac{\partial}{\partial\!\(m_f^2\)}\frac{n_F}{E_k}\(1-\frac{p_0}{P\cdot K}\) - \frac{n_F}{2E_k^3}\frac{p_0}{P\cdot K}\right\}\nn\\
&-&\sum_f\frac{ig^2(q_fB)^2p_0p_3}{3p_{\perp}^2}\!\int\!\frac{d^4K}{(2\pi)^4}\left[-\frac{\partial^2}{\partial\!\(m_f^2\)^2}+k^2(1-c^2)\frac{\partial^3}{\partial\!\(m_f^2\)^3}\right]\frac{k^0kc}{(K^2-m_f^2)(Q^2-m_f^2)}\nn\\
&=& F_1 + F_2,
\eea
where
\bea
F_1&=& -\sum_f\frac{g^2(q_fB)^2p^2}{\pi^2p_{\perp}^2}\Bigg[-g_k\left\{-\frac{p_0^2p_3^2}{3p^4}-\frac{A_0}{4}+\(\frac{3}{2}+\frac{p_0^2p_3^2}{p^4}\)A_2-\frac{5}{4}A_4\right\}\nn\\
&& +\(\frac{\pi}{32m_fT}-\frac{1}{8m_f^2}\)\left\{\frac{A_0}{4}-\(\frac{3}{2}+\frac{p_0^2p_3^2}{p^4}\)A_2+\frac{5}{4}A_4\right\}\nn\\
&& -f_k\frac{p_0^2p_3^2}{p^4}\(\frac{14}{3}-5A_0+A_2\)+\frac{p_0^2p_3^2}{p^4}\frac{8T-\pi m_f}{128Tm_f^2}\(5A_0-A_2\)\Bigg],  \label{f1}
\eea
\bea
F_2 &=&-\sum_f\frac{2ig^2(q_fB)^2p^2}{p_{\perp}^2} \frac{2}{3}\frac{p^0p^3}{p^2}\int\frac{d^4K}{(2\pi)^4}\left(-\frac{\partial^2}{\partial\!\(m_f^2\)^2}+k^2\(1-c^2\)\frac{\partial^3}{\partial\!\(m_f^2\)^3}\right)\nn\\
&&\times\ \frac{k_0kc}{(K^2-m_f^2)(Q^2-m_f^2)}\nn\\
&=& -\sum_f\frac{g^2(q_fB)^2}{6\pi^2m_fT}\frac{p^0p^3}{p_{\perp}^2}\frac{1}{1+\cosh{\frac{m_f}{T}}}\(\frac{3 A_1}{2}-A_3\). \label{f2}
\eea
\subsection{Calculation of the form factor \texorpdfstring{$a_2$}{a2} }
\label{a_eb_2}
\bea
2a_2&=&N^{\mn}(\delta \Pi^a_{\mn}+2\delta \Pi^b_{\mn})\nn\\
&=&\sum_f\frac{ig^2(q_fB)^2}{2} \int \frac{d^4K}{(2\pi)^4}\frac{N^{\mn}U_{\mn}}{(K^2-m_f^2)^2(Q^2-m_f^2)^2}+4i(e^2B)^2 \int \frac{d^4K}{(2\pi)^4}\bigg[\frac{N^{\mn}X_{\mn}}{(K^2-m_f^2)^3(Q^2-m_f^2)}\nn\\
&&-\frac{(k_0^2-k_3^2-m_f^2)N^{\mn}W_{\mn}}{(K^2-m_f^2)^4(Q^2-m_f^2)}\bigg]\nn\\
&=&\sum_f\frac{4ig^2(q_fB)^2}{\sqrt{\bar u^2}\sqrt{\bar n^2}} \int \frac{d^4K}{(2\pi)^4} \frac{p_0p_3}{p^2}\bigg[(-1+c^2)\frac{\partial}{\partial (m_f^2)}-\frac{1}{6}(5-3c^2)m_f^2\frac{\partial^2}{\partial (m_f^2)^2}\bigg]\nn\\
&&\times\frac{1}{(K^2-m_f^2)(Q^2-m_f^2)}+\frac{16i(e^2B)^2}{\sqrt{\bar u^2}\sqrt{\bar n^2}} \int \frac{d^4K}{(2\pi)^4} \bigg[\frac{1}{6}\frac{\partial^2}{\partial (m_f^2)^2}-\frac{k^2(1-c^2)}{3}\frac{\partial^3}{\partial (m_f^2)^3}\bigg]\nn\\
&&\times\frac{k_0k c}{(K^2-m_f^2)(Q^2-m_f^2)}\nn\\
&=&\sum_f\frac{4g^2(q_fB)^2}{\sqrt{\bar u^2}\sqrt{\bar n^2}} \int \frac{k^2 dk}{2\pi^2} \int \frac{d\Omega}{4\pi}\frac{p_0p_3}{p^2}\bigg[(-1+c^2)\frac{\partial}{\partial (m_f^2)}-\frac{1}{6}(5-3c^2)m_f^2\frac{\partial^2}{\partial (m_f^2)^2}\bigg]\nn\\
&&\times\Bigg\{-\frac{\partial}{\partial\!\(m_f^2\)}\frac{n_F}{E_k}\(1-\frac{p_0}{P\cdot \hat K}\) +\frac{n_F}{2E_k^3}\frac{p_0}{P\cdot \hat K}\Bigg\}-\frac{16(e^2B)^2}{\sqrt{\bar u^2}\sqrt{\bar n^2}}\int \frac{k^2 dk}{2\pi^2}\int \frac{d\Omega}{4\pi}\nn\\
&&\times \bigg[\frac{k c}{6}\frac{\partial^2}{\partial (m_f^2)^2}-\frac{k^3(c-c^3)}{3}\frac{\partial^3}{\partial (m_f^2)^3}\bigg]\frac{\partial n_F(E_k)}{\partial(m_f^2)}\bigg(1-\frac{p_0}{P\cdot \hat K}\bigg)\nn\\
&=& G_1 +G_2,
\eea
where
\bea
G_1&=&\sum_f\frac{4g^2(q_fB)^2}{\sqrt{\bar u^2}\sqrt{\bar n^2}} \int \frac{k^2 dk}{2\pi^2} \int \frac{d\Omega}{4\pi}\frac{p_0p_3}{p^2}\bigg[(-1+c^2)\frac{\partial}{\partial (m_f^2)}-\frac{1}{6}(5-3c^2)m_f^2\frac{\partial^2}{\partial (m_f^2)^2}\bigg]\nn\\
&&\times\Bigg\{-\frac{\partial}{\partial\!\(m_f^2\)}\frac{n_F}{E_k}\(1-\frac{p_0}{P\cdot \hat K}\) + \frac{n_F}{2E_k^3}\frac{p_0}{P\cdot \hat K}\Bigg\}\nn\\
&=&\sum_f\frac{4g^2(q_fB)^2}{\sqrt{\bar u^2}\sqrt{\bar n^2}} \int \frac{k^2 dk}{2\pi^2} \Bigg[\frac{p_0p_3}{p^2}\Bigg\{\bigg(\frac{2}{3}-A_0+A_2\bigg)\frac{\partial^2}{\partial ( m_f^2)^2}+\bigg(\frac{2}{3}-\frac{5A_0}{6}+\frac{A_2}{2}\bigg)m_f^2\frac{\partial^3}{\partial ( m_f^2)^3}\Bigg\}\nn\\
&&\times\frac{n_F}{E_k}+\Bigg\{\bigg(-A_0+A_2\bigg)\frac{\partial}{\partial ( m_f^2)}-\frac{1}{6}\bigg(5A_0-3A_2\bigg)m_f^2\frac{\partial^2}{\partial ( m_f^2)^2}\Bigg\}\frac{n_F}{2E_k^3}\Bigg]\nn\\
&=&\sum_f\frac{4g^2(q_fB)^2}{2\pi^2\sqrt{\bar u^2}\sqrt{\bar n^2}}\Bigg[\frac{p_0p_3}{p^2}\Bigg\{\bigg(\frac{2}{3}-A_0+A_2\bigg)g_k+\bigg(\frac{4}{3}-\frac{5A_0}{3}+A_2\bigg)f_k\Bigg\}\nn\\
&&+\Bigg\{\bigg(-A_0+A_2\bigg)\frac{\pi m_f-4T}{32 T m_f^2}-\frac{1}{6}\bigg(5A_0-3A_2\bigg)\frac{8T-\pi m_f}{64 T m_f^2}\Bigg\}\Bigg].\label{G1}
\eea
\bea
G_2&=&-\sum_f\frac{8g^2(q_fB)^2}{\sqrt{\bar u^2}\sqrt{\bar n^2}}\int \frac{k^2 dk}{2\pi^2}\int \frac{d\Omega}{4\pi} \bigg[\frac{k c}{6}\frac{\partial^2}{\partial (m_f^2)^2}-\frac{k^3(c-c^3)}{3}\frac{\partial^3}{\partial (m_f^2)^3}\bigg]\frac{\partial n_F(E_k)}{\partial(m_f^2)}\bigg(1-\frac{p_0}{P\cdot \hat K}\bigg)\nn\\
&=&\sum_f\frac{g^2(q_fB)^2}{\sqrt{\bar u^2}\sqrt{\bar n^2}6\pi^2 m_f T\big(1+\cosh{\frac{m_f}{T}}\big)}\bigg(-5A_1+4A_3\bigg).\label{G2}
\eea



\end{document}